\documentclass[journal=jpcl,manuscript=article]{achemso}
    \usepackage{achemso}
    \setkeys{acs}{usetitle = true}
    \usepackage[version=3]{mhchem} 
    \usepackage[T1]{fontenc}       
    \usepackage{color,soul}
    \usepackage{amsmath,amssymb}
    \usepackage{subfig}
    \usepackage{natbib}
\usepackage{array}
\usepackage{multirow}
\usepackage{makecell}

 \author{Divyanshi Tyagi}
 \email{divyanshi.tyagi@physics.iitd.ac.in[DT]}
 \author{Saswata Bhattacharya}
 \email{saswata@physics.iitd.ac.in[SB]}
 \affiliation[Indian Institute of Technology Delhi]
 {Department of Physics, Indian Institute of Technology Delhi, Hauz Khas, New Delhi 110016, India}

    \title[An \textsf{achemso} demo]
    {Ferroelectric Control of Spin Textures in Layered Hybrid Perovskites}
 

\begin{tocentry}
\centering
\begin{figure}[H]%
	\includegraphics[width=0.7\columnwidth,clip]{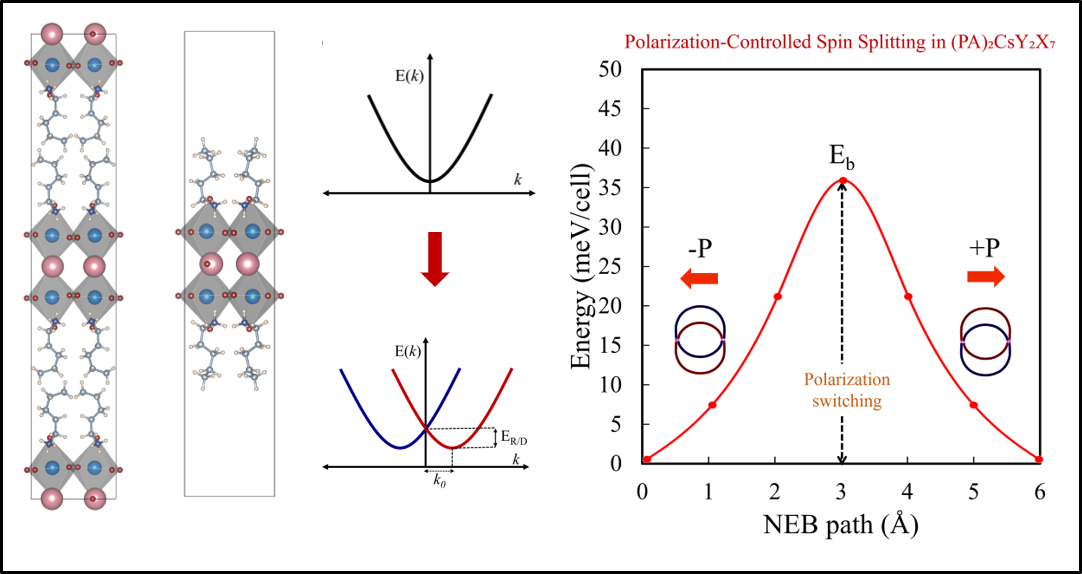}
	\end{figure}
\end{tocentry}

\begin{document}
 
\begin{abstract}
Hybrid organic--inorganic perovskites with broken inversion symmetry provide a fertile ground for uncovering coupled spin-orbit and ferroelectric phenomena. Here, we investigate the layered family (PA)$_2$CsY$_2$X$_7$ (Y = Pb, Sn; X = I, Br) using density functional theory, Berry-phase polarization analysis, and effective $\boldsymbol{k \cdot p}$
 modeling. Across all four members, we find indirect bandgaps with extrema near $\Gamma$, sizable spin splittings at both band edges, and robust in-plane ferroelectric polarization that stabilizes out-of-plane persistent spin textures (PSTs). Crucially, polarization reversal switches the spin orientation, enabling electrical control of PSTs and thereby non-volatile manipulation of spin states. These results establish (PA)$_2$CsY$_2$X$_7$ as a versatile materials platform where compositional design and ferroelectric switching jointly enable spintronic functionality.
\end{abstract}

In recent years, hybrid organic--inorganic perovskites (HOIPs) have gained significant attention owing to their tunable chemical structures and exceptional optoelectronic properties, with demonstrated applications in solar cells,\cite{park2020scalable,wang2021prospects,wang2022spacer} light-emitting diodes (LEDs),\cite{tsai2020critical,khan2022high,liu2021metal} lasers,\cite{qin2020stable,zhang2021halide} photodetectors,\cite{zhang2021lead,li2021supersaturation} and photocatalysis.\cite{dave2020recent} These materials generally adopt an ABX$_3$ perovskite structure, where \textbf{A} is an organic or inorganic cation (e.g., MA = methylammonium, FA = formamidinium, or Cs$^+$), \textbf{B} is typically a divalent metal cation (e.g., Ge$^{2+}$, Sn$^{2+}$, or Pb$^{2+}$), and \textbf{X} is a halide anion (Cl$^-$, Br$^-$, or I$^-$).\cite{protesescu2015nanocrystals} The resulting three-dimensional (3D) framework consists of corner-sharing BX$_6$ octahedra. While Pb-based HOIPs have shown excellent optoelectronic performance, their toxicity raises concerns for large-scale deployment. To address this issue, both lead-free variants and low-dimensional HOIPs have emerged as promising alternatives.

Structural engineering through the incorporation of bulky organic cations has enabled the realization of quasi-two-dimensional (2D) perovskites. These layered materials, consisting of alternating organic and inorganic sheets, offer enhanced environmental stability and strong quantum confinement. Their reduced symmetry also breaks inversion symmetry, thereby allowing access to novel spin-dependent phenomena.

One of the most compelling aspects of HOIPs---especially those incorporating heavy elements such as Sn or Pb---is the presence of strong spin--orbit coupling (SOC). In non-centrosymmetric crystals, SOC manifests as an effective momentum-dependent magnetic field $\boldsymbol{k \cdot p}$, lifting spin degeneracy and generating Rashba- or Dresselhaus-type spin splittings.\cite{rashba1960spin,dresselhaus1955spin,marchenko2012giant} The resulting spin textures, defined by momentum-locked spin orientations, can support persistent spin textures (PSTs)---long-lived spin configurations resistant to dephasing. While PSTs have been mainly studied in semiconductor quantum wells and oxide heterostructures,\cite{sasaki2014direct,kohda2012gate,walser2012direct,koralek2009emergence,tao2018persistent,autieri2019persistent} 2D HOIPs represent an unexplored platform where large SOC and intrinsic polar asymmetry can naturally stabilize such states.

In this work we systematically investigate the layered family (PA)$_2$CsY$_2$X$_7$ (Y = Pb, Sn; X = I, Br) using first-principles density functional theory (DFT) calculations combined with a symmetry-adapted $\boldsymbol{k \cdot p}$ model. While calculations were performed for all four variants, we present detailed results for the representative tin--bromide compound (PA)$_2$CsSn$_2$Br$_7$ (both bulk and monolayer forms) and summarize the Pb- and I-containing analogues in the Supporting Information(SI).  

Our analysis shows that a sizable in-plane ferroelectric (FE) polarization -- hereafter taken along the $z$-axis within the perovskite layer -- stabilizes a unidirectional, out-of-plane spin texture at the band edges. The spin texture is dominated by the $S_x$ component in both valence and conduction bands and is accompanied by pronounced SOC-driven spin splittings. Importantly, these spin textures are fully reversible under polarization switching, thereby coupling FE order to spin degrees of freedom. \cite{tao2017reversible}  

The microscopic origin of the unidirectional spin textures is analyzed via a symmetry-constrained $\boldsymbol{k \cdot p}$ Hamiltonian that captures the interplay between in-plane FE distortions and the crystal's $C_{2v}$ symmetry. We further propose that applying an external in-plane electric field opposite to the intrinsic polarization can reverse the orientation of the out-of-plane spin textures. These results establish layered (PA)$_2$CsY$_2$X$_7$ compounds as promising platforms for non-volatile electrical control of spin states in spintronic devices.

\begin{figure}[h!]
\centering
\includegraphics[width=0.6\textwidth]{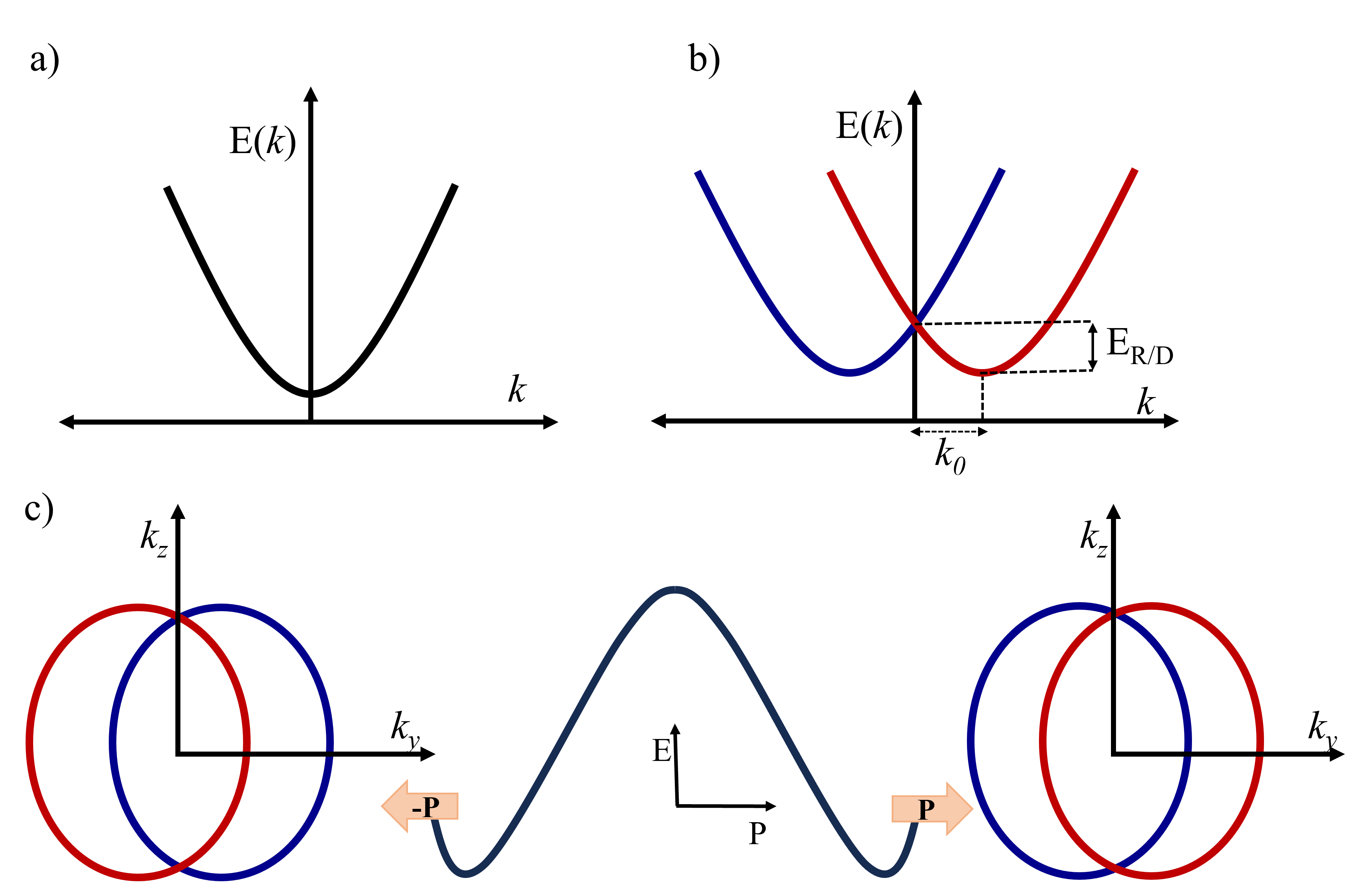}
\caption{Schematic representation of spin--orbit-induced band splitting and polarization-dependent spin texture reversal.  
(a) Parabolic band structure in the absence of spin--orbit coupling (SOC), showing spin-degenerate bands.  
(b) SOC-induced spin splitting with characteristic energy offset $E_{\mathrm{R/D}}$ and momentum shift $k_0$, illustrating the emergence of spin-split bands with out-of-plane spin orientation.  
(c) Depiction of spin textures in the $k_y$--$k_z$ plane under opposite ferroelectric polarizations ($\pm\mathbf{P}$). The red and blue loops correspond to spin textures with opposite $S_x$ orientations, indicating the formation of persistent spin textures (PSTs).  
Switching the in-plane polarization $\mathbf{P}$ reverses the spin texture, highlighting the intimate coupling between ferroelectricity and spin degrees of freedom.}
\label{fig:sche}
\end{figure}

We focus here on the representative tin--bromide compound (PA)$_2$CsSn$_2$Br$_7$ in both bulk and monolayer forms; results for the other family members, along with convergence tests and additional structural variants, are provided in the SI. 

\begin{figure}[h!]
\centering
\includegraphics[width=0.45\textwidth]{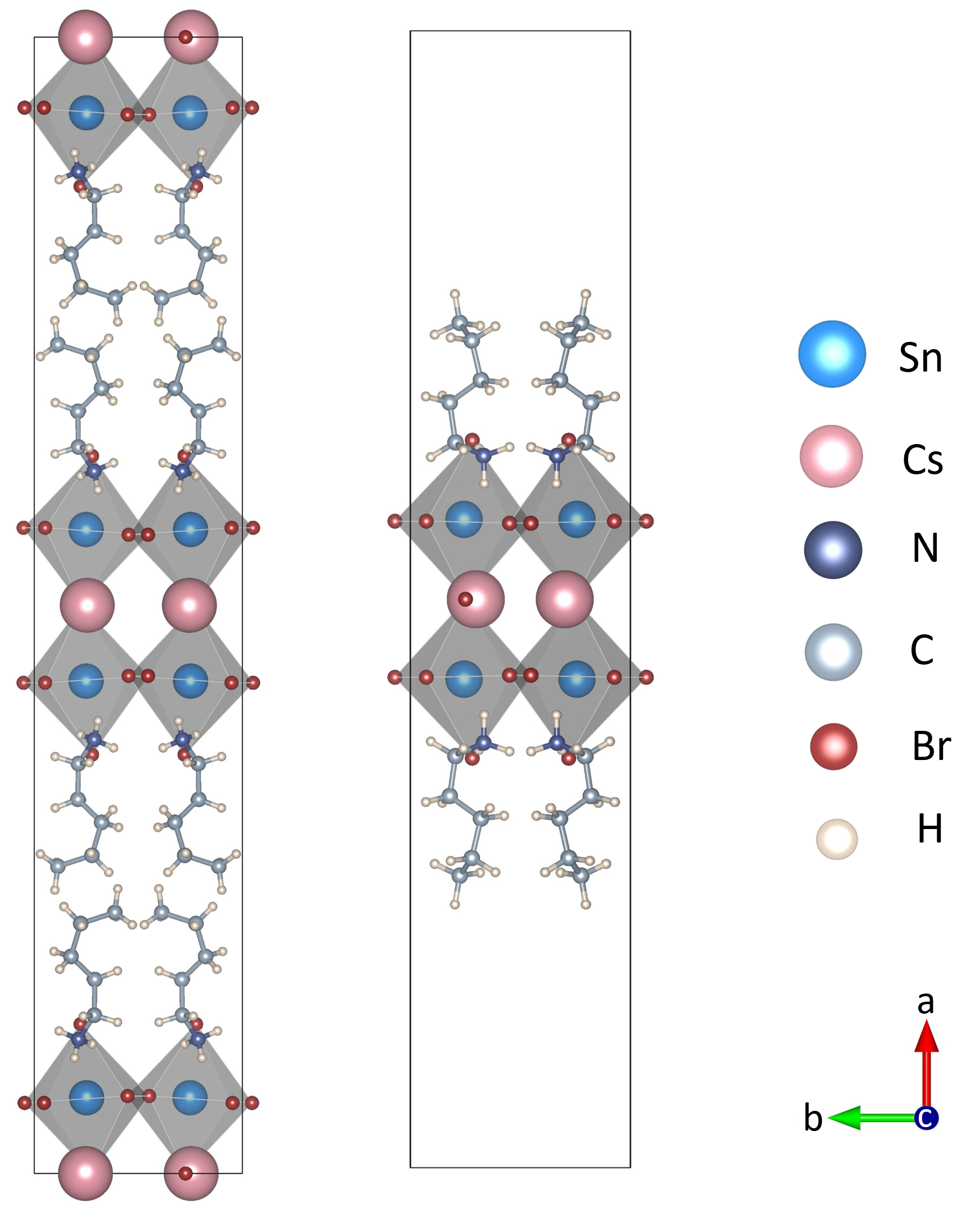}
\caption{Side view of the relaxed crystal structures of bulk (left) and monolayer (right) (PA)$_2$CsSn$_2$Br$_7$. Sn-centered \(\mathrm{SnBr}_6\) octahedra form the inorganic sheets separated by organic \(\mathrm{PA}^+\) cations.}
\label{fig:structure}
\end{figure}

As shown in Fig.~\ref{fig:structure}, the lattice is composed of two essential building blocks: (i) inorganic sheets formed by corner-sharing \(\mathrm{SnBr}_6\) octahedra extending along the $b$- and $c$-axes, and (ii) organic cations \(\mathrm{[C_5H_{11}NH_3]^+}\) and \(\mathrm{Cs^+}\), whose orientations govern the spontaneous polarization. At ambient conditions, the bulk structure crystallizes in the polar orthorhombic phase \textit{Cmc}2$_1$ with lattice parameters $a = 45.14$~\AA, $b = 8.25$~\AA, and $c = 8.22$~\AA. The FE polarization points along the in-plane $z$-axis, while mirror symmetry forbids contributions along $x$, and dipoles cancel along $y$.

The adjacent monolayers are connected by weak van der Waals forces,\cite{wei2019regulating} which makes exfoliation experimentally feasible. In the monolayer (Fig.~\ref{fig:structure}, right), two \(\mathrm{PA^+}\) and one \(\mathrm{Cs^+}\) ions are preserved to maintain charge neutrality. After structural relaxation, the system stabilizes in the polar orthorhombic \textit{Pmc2$_1$} phase with lattice constants $a = 42.62$~\AA, $b = 8.25$~\AA, and $c = 8.22$~\AA. The polarization orientation remains along the in-plane $z$-axis. Both bulk and monolayer are polar and non-centrosymmetric, enabling SOC-induced spin splitting and electrically switchable spin textures.

\begin{figure}[t]
\centering
\includegraphics[width=0.6\textwidth]{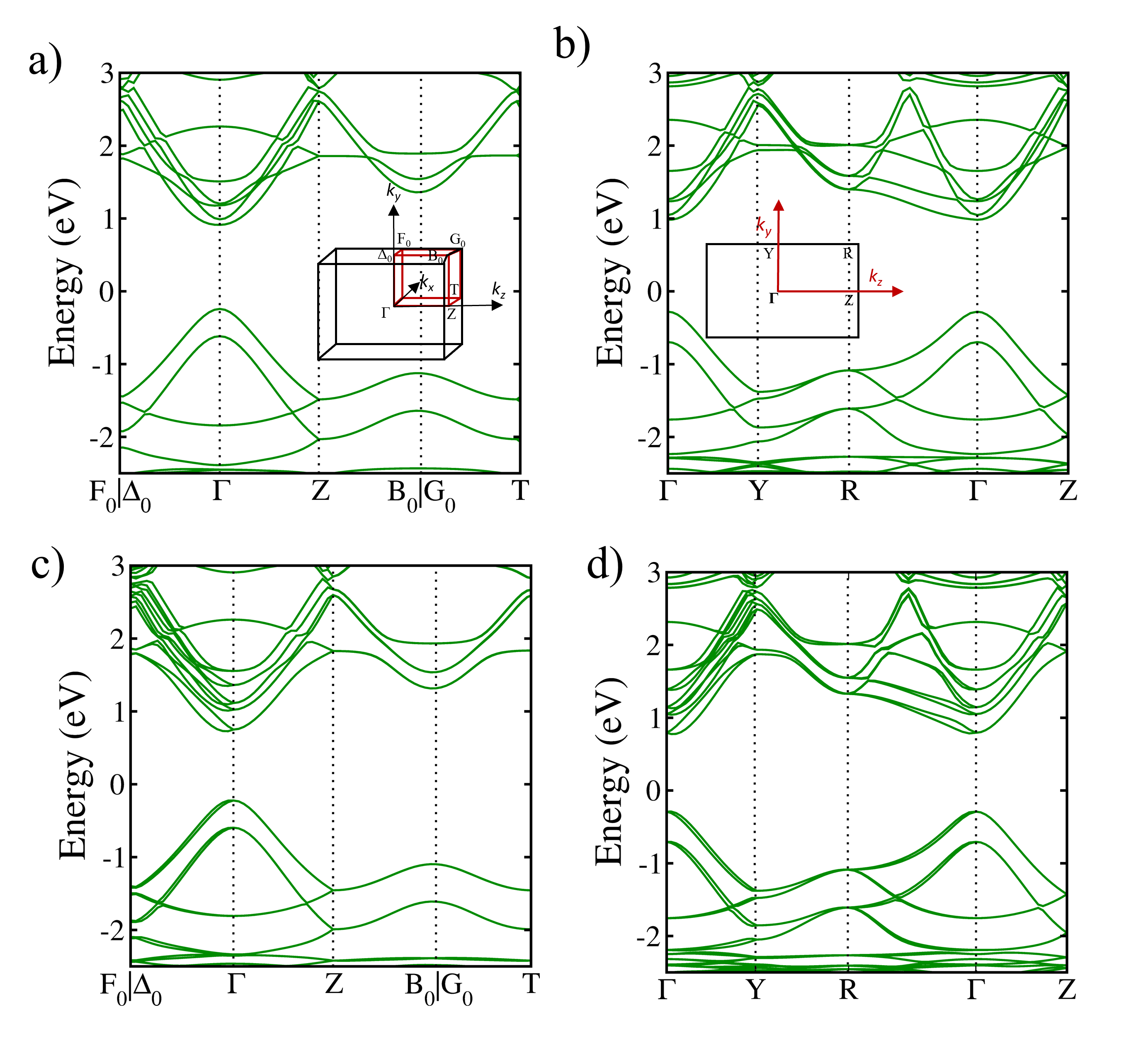}
\caption{Electronic band structures of bulk (a, c) and monolayer (b, d) (PA)$_2$CsSn$_2$Br$_7$. 
Panels (a, b) show results without spin--orbit coupling (SOC), while (c, d) include SOC. 
Insets illustrate the corresponding high-symmetry $k$-paths in the first Brillouin zone with coordinate axes.}
\label{fig:bands}
\end{figure}

Also the bulk and monolayer (PA)$_2$CsSn$_2$Br$_7$ both are moderate-gap semiconductors. The bulk exhibits a band gap of 1.15~eV, while the monolayer shows a slightly larger value of 1.26~eV. As illustrated in Fig.~\ref{fig:bands}, their dispersions are similar: the valence band maximum (VBM) lies at the $\Gamma$ point, and the conduction band minimum (CBM) is slightly displaced from $\Gamma$. Thus, although the systems are indirect-gap semiconductors, their optical responses are expected to be comparable to those of direct-gap materials.

The inclusion of SOC reduces the gaps to 0.95~eV (bulk) and 1.06~eV (monolayer), primarily due to SOC-induced band shifts near $\Gamma$. Quantitatively, the VBM splitting is 0.41~eV$\cdot$\AA \ and the CBM splitting is 1.01~eV$\cdot$\AA \ in the bulk. In the monolayer, these values increase to 0.69~eV$\cdot$\AA \ (VBM) and 1.60~eV$\cdot$\AA \ (CBM), reflecting enhanced SOC effects in reduced dimensionality.

From symmetry considerations, the $C_2$ rotation axis along [001] forbids spin splitting along the $\Gamma$--$Z$ direction, whereas finite splitting is allowed along $\Gamma$--$Y$, consistent with the band structure. The pronounced SOC-induced splittings, particularly in the monolayer, underscore the strong coupling between electronic and spin degrees of freedom in this system.

This compound is derived from a reported lead-based analogue, where Pb was replaced by Sn while retaining the same perovskite framework~\cite{guo2023electrically}. Among the four variants studied [ (PA)$_2$CsSn$_2$Br$_7$, (PA)$_2$CsSn$_2$I$_7$, (PA)$_2$CsPb$_2$Br$_7$, and (PA)$_2$CsPb$_2$I$_7$; see SI for details ], the tin--bromide compound exhibits the largest SOC-induced band splittings in both valence and conduction bands, while also offering a lead-free (and therefore less toxic) alternative to the Pb-based analogues.

Taken together, these results establish (PA)$_2$CsSn$_2$Br$_7$ as a promising tunable platform for realizing electrically switchable spin textures, with potential applications in spintronic and non-volatile logic devices.

The spin textures around the $\Gamma$ point can be understood in terms of an effective $\boldsymbol{k \cdot p}$ Hamiltonian, which can be deduced from symmetry considerations. 
Throughout this section $x$ denotes the out-of-plane direction (normal to the layer), while $y$ and $z$ are in-plane. The ferroelectric polarization is along $z$, and the observed spin textures are dominated by the out-of-plane $S_x$ component.

\begin{table*}
\centering
\normalsize
\caption{Transformations of $(\sigma_x,\sigma_y,\sigma_z)$ and $(k_x,k_y,k_z)$ under the action of the generators of the $C_{2v}$ point group and the time-reversal operator $T$. The first row lists the point-group operations and their corresponding symmetries. Since the generators fully define the group structure, only these generators, along with the time-reversal operation $T = i\sigma_y K$ (where $K$ is the complex-conjugation operator), are used in constructing the $\boldsymbol{k \cdot p}$. The last row shows the terms that remain invariant under the point-group operations. Terms up to cubic order in $k$ are included, as higher-order contributions are negligible.}
\vspace{0.5em}
\setlength{\tabcolsep}{6pt}
\renewcommand{\arraystretch}{2.5}
\begin{tabular}{|>{\bfseries\arraybackslash}c|c|c|c|}
\hline
Operations & $(k_x,k_y,k_z)$ & $(\sigma_x,\sigma_y,\sigma_z)$ & Invariants \\
\hline
$C_{2z}=e^{-i\pi/2\,\sigma_z}$ &
$(-k_x,\,-k_y,\,k_z)$ &
$(-\sigma_x,\,-\sigma_y,\,\sigma_z)$ &
\makecell[l]{%
  $k_i^m k_x\sigma_x,\;k_i^m k_x\sigma_y,k_i^m k_y\sigma_x,$\\
  $k_i^m k_y\sigma_y,\;k_i^m k_z\sigma_z \;(i=x,y,z;\;m=0,2)$%
} \\
\hline
$M_{yz}=i\sigma_x$ &
$(-k_x,\;k_y,\;k_z)$ &
$(\sigma_x,\,-\sigma_y,\,-\sigma_z)$ &
\makecell[l]{%
  $k_i^m k_x\sigma_y,\;k_i^m k_y\sigma_x,$\\
  $k_i^m k_x\sigma_z,\:k_i^m k_z\sigma_x\;(i=x,y,z;\;m=0,2)$%
} \\
\hline
$M_{xz}=i\sigma_y$ &
$(k_x,\,-k_y,\;k_z)$ &
$(-\sigma_x,\;\sigma_y,\,-\sigma_z)$ &
\makecell[l]{%
  $k_i^m k_x\sigma_y,\;k_i^m k_y\sigma_x,$\\
  $k_i^m k_y\sigma_z,\;k_i^m k_z\sigma_y\;(i=x,y,z;\;m=0,2)$%
} \\
\hline
$T=i\sigma_yK$ &
$(-k_x,\,-k_y,\,-k_z)$ &
$(-\sigma_x,\,-\sigma_y,\,-\sigma_z)$ &
$k_i\,\sigma_j\;(i,j=x,y,z)$ \\
\hline
\end{tabular}
\label{table1}
\end{table*}

\begin{figure*}
\centering
\includegraphics[width=1\textwidth]{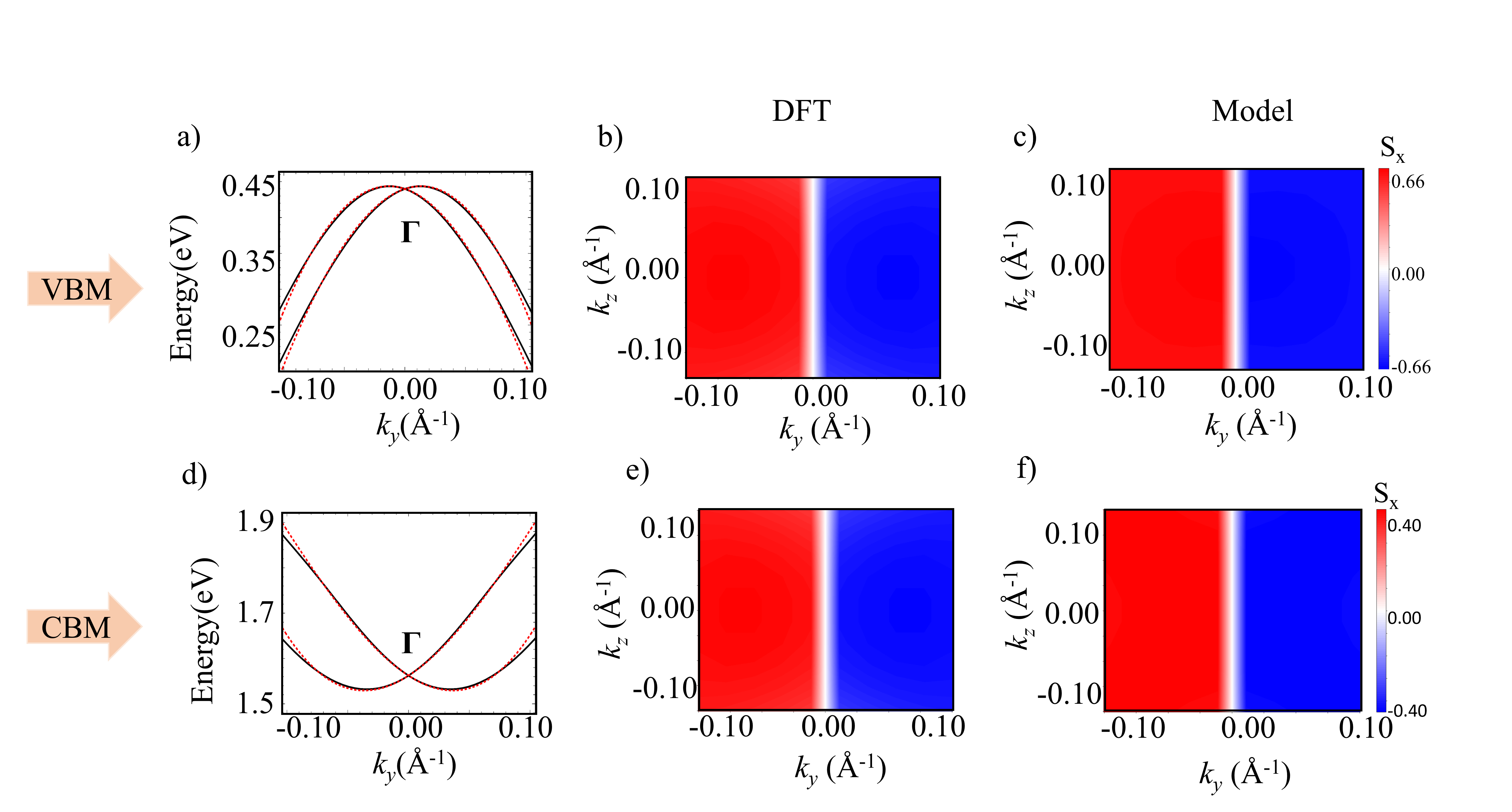}
\caption{Bulk (PA)$_2$CsSn$_2$Br$_7$: unidirectional out-of-plane spin textures near $\Gamma$. 
Top row = VBM, bottom row = CBM. 
(a,d) DFT bands (black) and $\boldsymbol{k \cdot p}$ fits (red dashed); 
(b,e) DFT $S_x$ on constant-energy contours; 
(c,f) $S_x$ from the two-band $\boldsymbol{k \cdot p}$ model. 
Spin textures are dominated by the out-of-plane component $S_x$ (colorbar ranges differ between rows).}

\label{fig:bulk_pst}
\end{figure*}

\begin{figure*}
\centering
\includegraphics[width=1\textwidth]{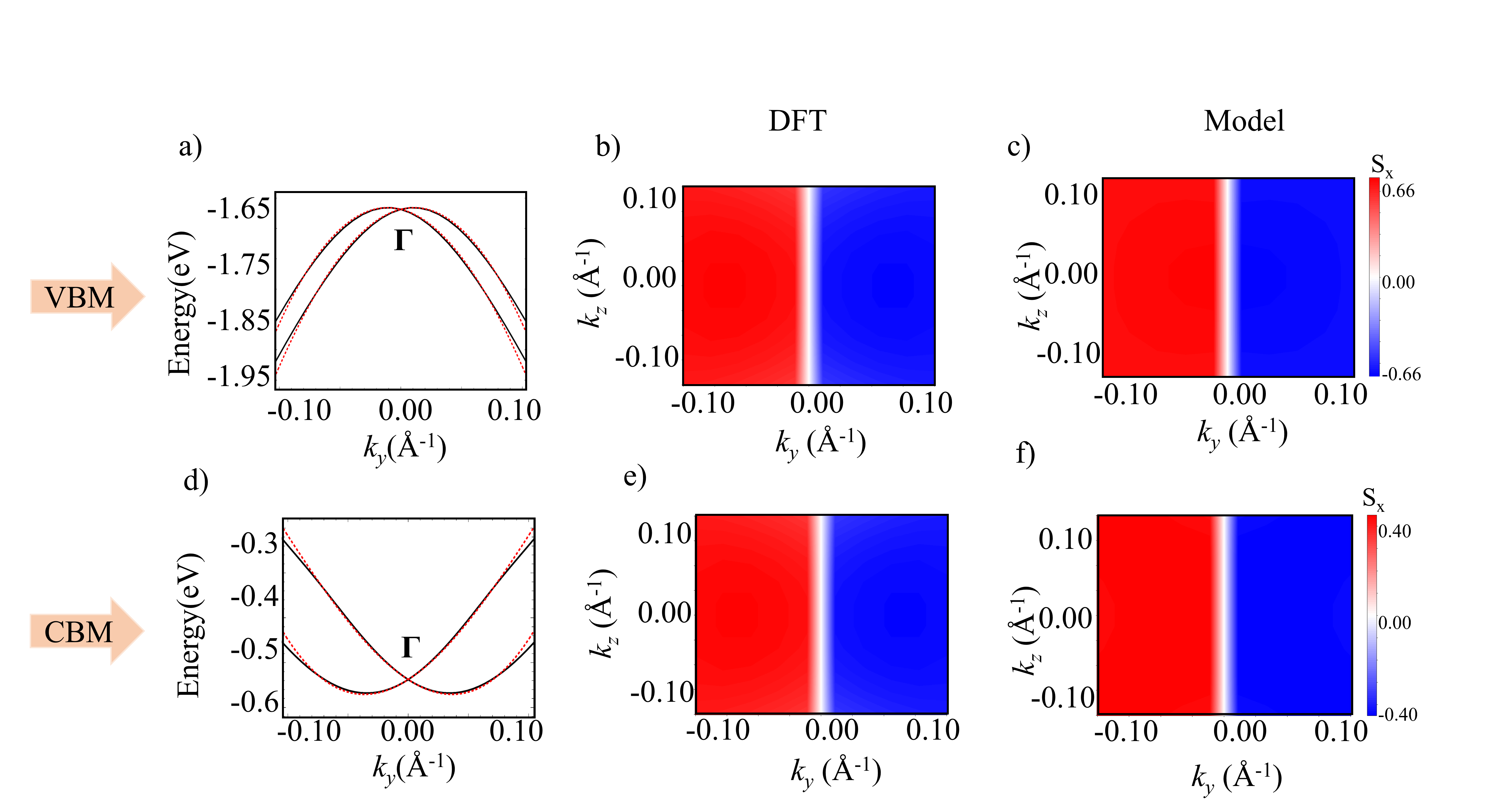}
\caption{Monolayer (PA)$_2$CsSn$_2$Br$_7$: unidirectional out-of-plane spin textures near $\Gamma$. 
Top row = VBM, bottom row = CBM. 
(a,d) DFT bands (black) and $\boldsymbol{k \cdot p}$ fits (red dashed); 
(b,e) DFT $S_x$ on constant-energy contours; 
(c,f) $S_x$ from the two-band $\boldsymbol{k \cdot p}$ model. 
Spin textures are dominated by the out-of-plane component $S_x$ (colorbar ranges differ between rows).}

\label{fig:monopst}
\end{figure*}

Figures~\ref{fig:bulk_pst} and \ref{fig:monopst} show the band fitting by $\boldsymbol{k \cdot p}$ model and the associated spin textures around the $\Gamma$ point, for both the systems bulk and monolayer respectively. The spin-split bands are predominantly linear in $\mathbf{k}$. For deeper insight, the band dispersion can be deduced by identifying all symmetry-allowed terms such that the Hamiltonian satisfies the symmetry condition:
\[
H(\mathbf{k}) = \mathcal{O}^\dagger H(\mathbf{k}) \mathcal{O},
\]
where $\mathcal{O}$ is a symmetry operation belonging to the group of the wave vector $G$ associated with the high-symmetry point, along with time-reversal symmetry $T$. The invariant Hamiltonian must satisfy the relation:
\[
H_G(\mathbf{k}) = D(\mathcal{O}) H(\mathcal{O}^{-1} \mathbf{k}) D^{-1}(\mathcal{O}), \quad \forall \, \mathcal{O} \in G, T,
\]
where $D(\mathcal{O})$ is the matrix representation of $\mathcal{O}$ in the point group $G$. The $\boldsymbol{k \cdot p}$ Hamiltonian is derived using the method of invariants by considering the little group of the $\Gamma$ point to be $C_{2v}$, which consists of mirror planes $M_{xz}$ and $M_{yz}$, in addition to the identity operation ($E$). Although the bulk (\textit{Cmc}2$_1$) and monolayer (\textit{Pmc}2$_1$) contain non-symmorphic operations such as screw axes and glide planes, these act trivially at the $\Gamma$ point, reducing both to the same symmorphic point group $mm2$. As a result, the PST observed near $\Gamma$ is governed solely by the point-group symmetry. 

The $\boldsymbol{k \cdot p}$ Hamiltonian around the $\Gamma$ point, following the transformation rules listed in Table~\ref{table1}, is given
~\cite{kumar2023exploring}
\begin{equation}
H_{C_{2v}}(\mathbf{k}) = H_0(\mathbf{k}) + \alpha(k) k_x \sigma_y + \beta(k) k_y \sigma_x,
\label{eq:HC2v}
\end{equation}
where $H_0(\mathbf{k})$ is the Hamiltonian that describes the band dispersion, and depends on the parameters $\kappa$ and $\delta$ as:
\begin{equation}
H_0(\mathbf{k}) = E_0 + \kappa k_x^2 + \delta k_y^2,
\label{eq:H0}
\end{equation}

\noindent
Here, $\kappa$ and $\delta$ are related to the effective masses ($m_x^*$, $m_y^*$) by the expressions
\[
|\kappa| = \frac{\hbar^2}{2 m_x^*}, \quad |\delta| = \frac{\hbar^2}{2 m_y^*},
\]
respectively. ${\sigma}$ denotes the vector of Pauli matrices that describe the spin degrees of freedom. The parameters $\alpha$ and $\beta$ are linear SOC splitting coefficients associated with the symmetry-allowed spin-momentum coupling terms.

The $k$-dependent spin-orbit coupling constants in $H_{C_{2v}}(\mathbf{k})$ are given by
\begin{align}
\alpha(\mathbf{k}) &= \alpha^{(1)} + \alpha^{(3)}k^2 + \gamma_\alpha(k_x^2 - k_y^2), \tag{6} \\
\beta(\mathbf{k}) &= \beta^{(1)} + \beta^{(3)}k^2 + \gamma_\beta(k_x^2 - k_y^2), \tag{7}
\end{align}
where $\alpha^{(3)}$ and $\beta^{(3)}$ represent the $k^2$-dependent renormalization terms for the linear spin-orbit coupling constants, while $\gamma_\alpha$ and $\gamma_\beta$ account for the $k$-cubic anisotropic interactions.

The eigenstates corresponding to Hamiltonian $H_{C_{2v}}(\mathbf{k})$ are
\begin{align}
\psi_{\mathbf{k}, \uparrow} &= \frac{e^{i\mathbf{k} \cdot \mathbf{r}}}{\sqrt{2\pi}} \begin{pmatrix} -\frac{\Lambda_x(\mathbf{k}) - i\Lambda_y(\mathbf{k})}{E_{SO}} \\ 1 \end{pmatrix}, \tag{8} \\
\psi_{\mathbf{k}, \downarrow} &= \frac{e^{i\mathbf{k} \cdot \mathbf{r}}}{\sqrt{2\pi}} \begin{pmatrix} \frac{\Lambda_x(\mathbf{k}) - i\Lambda_y(\mathbf{k})}{E_{SO}} \\ 1 \end{pmatrix}, \tag{9}
\end{align}
where
\[
\Lambda_x(\mathbf{k}) = \beta(\mathbf{k})k_y + \delta(k_y^3 + k_x^2k_y), \quad 
\Lambda_y(\mathbf{k}) = \alpha(\mathbf{k})k_x + \gamma(k_x^3 + k_xk_y^2),
\]
and
\[
E_{SO} = \sqrt{\Lambda_x(\mathbf{k})^2 + \Lambda_y(\mathbf{k})^2}.
\]

The corresponding eigenvalues are given as
\begin{equation}
E_{C_{2v}}(\mathbf{k}) = E_0(\mathbf{k}) \pm E_{SO}. \tag{10}
\end{equation}

The spin textures, determined by the expression $\mathbf{S}_\pm = \frac{\hbar}{2} \langle \psi_{\mathbf{k}, \downarrow} | {\sigma} | \psi_{\mathbf{k}, \downarrow} \rangle$, are given as
\begin{align}
\langle \sigma_x^\pm \rangle &= \pm \frac{\beta(\mathbf{k})k_y}{\sqrt{\alpha(\mathbf{k})^2k_x^2 + \beta(\mathbf{k})^2k_y^2}}, \tag{11} \\
\langle \sigma_y^\pm \rangle &= \pm \frac{\alpha(\mathbf{k})k_x}{\sqrt{\alpha(\mathbf{k})^2k_x^2 + \beta(\mathbf{k})^2k_y^2}}, \tag{12} \\
\langle \sigma_z^\pm \rangle &= \pm 0. \tag{13}
\end{align}
Since our 2D ferroelectric system lies on the $y$--$z$ plane, we obtain that the $k_x \sigma_y$ term of the $\mathcal{H}_{\mathrm{C2v}}(k)$ in Eq.~(1) naturally disappears; thus, Eq.~\eqref{eq:HC2v} simplifies to:
\begin{equation}
H_{C_{2v}}(\mathbf{k}) = H_0(\mathbf{k}) + \beta(k) k_y \sigma_x.
\end{equation}

\noindent
The spin textures, determined from the expectation value
\[
\mathbf{S}^{\pm} = \frac{\hbar}{2} \langle \Psi_{\mathbf{k}\uparrow, \downarrow} | {\sigma} | \Psi_{\mathbf{k}\uparrow, \downarrow} \rangle,
\]
are given by:
\begin{equation}
\mathbf{S}^{\pm} = \pm \frac{\hbar}{2}(1, 0, 0).
\label{eq:spintexture}
\end{equation}

\noindent
Equation~\eqref{eq:spintexture} clearly shows that only the out-of-plane spin component is present around the $\Gamma$ point. This is consistent with the spin texture obtained from DFT calculations, as shown in Fig.~\ref{fig:bulk_pst} and Fig.~\ref{fig:monopst}.

By fitting the DFT band dispersion and spin textures near the $\Gamma$ point, 
we extract the leading linear spin--orbit coefficients $\beta^{(1)}$ 
(in eV\,$\cdot$\,\AA) and the higher-order coefficient $\delta$ 
(in eV\,$\cdot$\,\AA$^3$). For the bulk system, the values are
\begin{align*}
\text{VBM:} \quad & \beta^{(1)} = 0.56~\text{eV}\cdot\text{\AA}, 
& \delta = -18.07~\text{eV}\cdot\text{\AA}^3, \\
\text{CBM:} \quad & \beta^{(1)} = 1.00~\text{eV}\cdot\text{\AA}, 
& \delta = -47.45~\text{eV}\cdot\text{\AA}^3.
\end{align*}
For the monolayer, the fitted values are
\begin{align*}
\text{VBM:} \quad & \beta^{(1)} = 0.67~\text{eV}\cdot\text{\AA}, 
& \delta = 36.65~\text{eV}\cdot\text{\AA}^3, \\
\text{CBM:} \quad & \beta^{(1)} = 1.38~\text{eV}\cdot\text{\AA}, 
& \delta = -56.05~\text{eV}\cdot\text{\AA}^3.
\end{align*}

The extracted $\beta^{(1)}$ coefficients indicate sizable linear spin splittings. 
For context, these values are significantly larger than those reported in related layered HOIPs, such as (BA)$_2$PbCl$_4$ (0.69~eV$\cdot$\AA{}) and (FPEA)$_2$PbI$_4$ (0.36~eV$\cdot$\AA{}).

\noindent

Furthermore, the effective SOC-induced magnetic field $\boldsymbol{\Omega}_\mathbf{k} = \alpha_D (k_y, 0, 0)$ is perpendicular to the 2D plane ($x = 0$), indicating that this system can host a PST when spin is injected. This leads to the expectation of a long spin lifetime.

\begin{figure}[h!]
\centering
\includegraphics[width=0.7\textwidth]{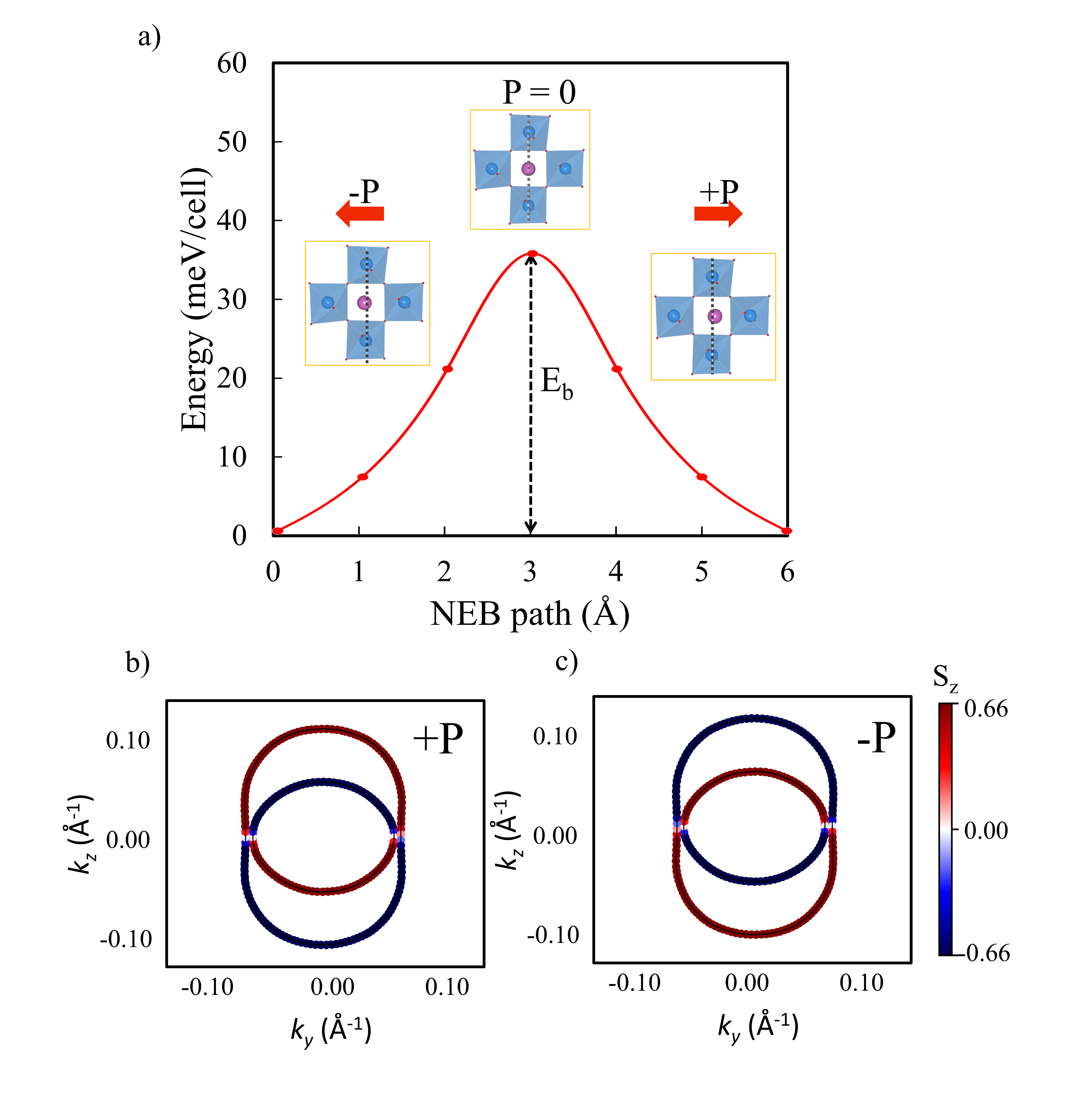}
\caption{%
(a) Climbing--image nudged elastic band (CI--NEB) calculation for the polarization switching process in bulk (PA)$_2$CsSn$_2$Br$_7$ perovskite. Two ferroelectric structures in the ground state with opposite directions of electric polarization are shown. $E_b$ is the activation barrier energy for the polarization switching process. Reversible in--plane spin textures calculated at constant energy $E = E_F + 1.5$eV with opposite spin polarization: (b) $-P$, (c) $+P$.%
}
\label{fig:neb}
\end{figure}

Lastly, we have explored the possibility of polarization switching in FE (PA)$_2$CsSn$_2$Br$_7$. To predict the feasibility of this phenomenon, we have analyzed the minimum energy pathway of the FE transition using the climbing-image NEB method. To estimate the theoretical FE polarization, we need to introduce a reference centrosymmetric structure related to the FE phase through complex atomic distortions. All detailed information regarding how to construct such a reference structure in complex organic--inorganic systems can be found in Refs.~\cite{hu2017dipole,hu2017ferroelectric,stroppa2015ferroelectric}. As shown in Figure~\ref{fig:neb}(a), the energy barrier $E_b$ for polarization switching is estimated to be 36~meV for bulk system, suggesting that switchable FE polarization switching is indeed plausible in this material. The polarization switching for monolayer is shown in the SI.

\begin{figure}[h!]
\centering
\includegraphics[width=0.9\textwidth]{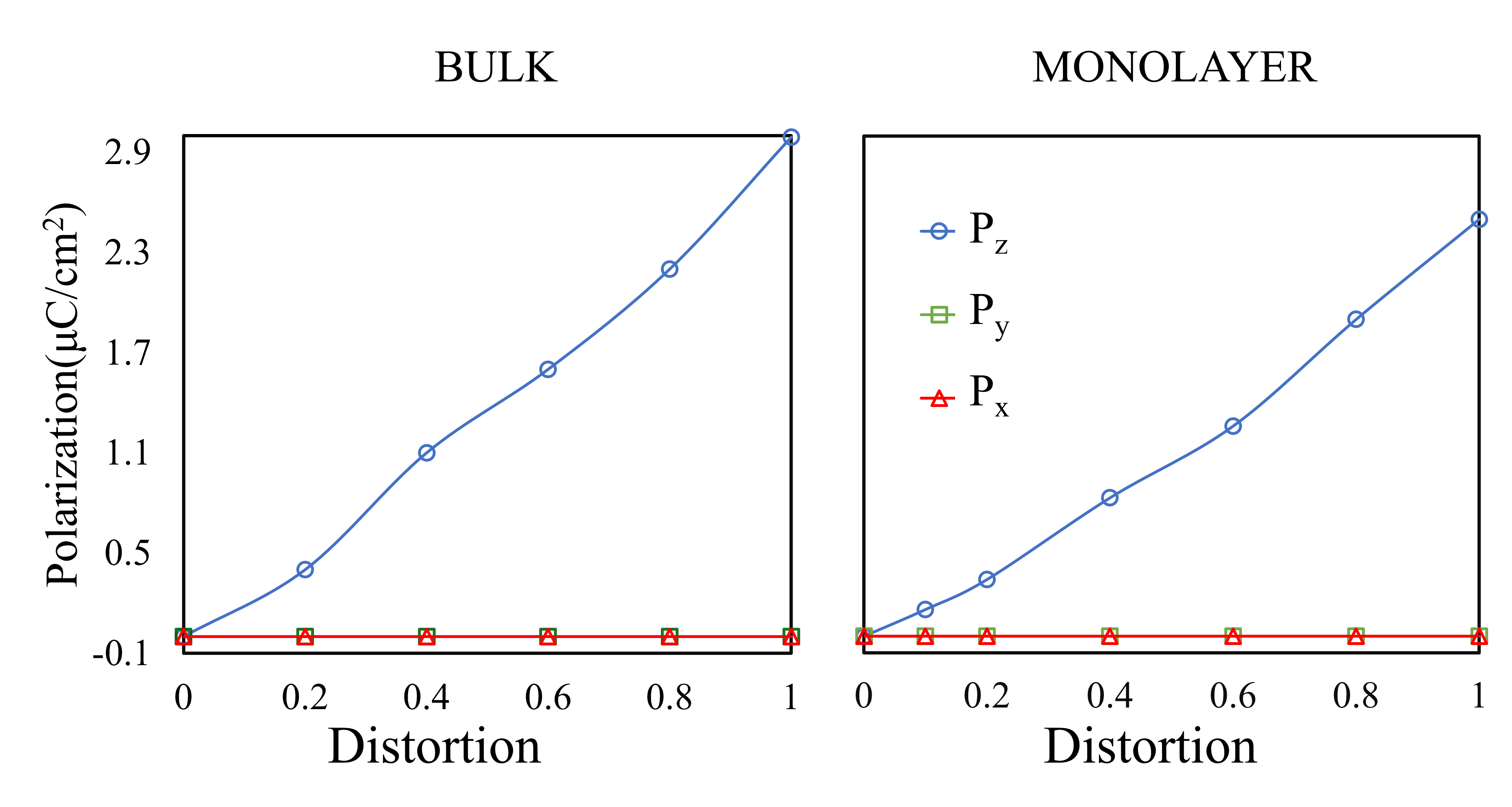}
\caption{Variation of the Berry-phase polarization of bulk (a) and monolayer (b) (PA)$_2$CsSn$_2$Br$_7$ as a function of the normalized polar distortion amplitude $\lambda$ between the centrosymmetric reference ($\lambda=0$) and the relaxed polar structure ($\lambda=1$).}
\label{fig:polar}
\end{figure}

We compute the electric polarization with the modern Berry-phase approach. The polarization curves in Fig.~\ref{fig:polar} are plotted as a function of a normalized distortion amplitude $\lambda$, where $\lambda=0$ denotes the chosen centrosymmetric reference structure and $\lambda=1$ corresponds to the fully relaxed polar structure; at each $\lambda$ the ionic positions are interpolated between these endpoints and the electronic contribution to $P$ is evaluated via the Berry-phase formula. Since the $z$-axis is the only polar axis in both structures, only the $P_z$ component contributes to the net polarization. The estimated Berry-phase polarization values are comparable for the bulk and monolayer systems--3.0 and 2.5~$\mu$C/cm$^2$, respectively. The small difference between these values indicates that the individual inorganic layers are effectively decoupled, in contrast to conventional 3D inorganic perovskites where the polarization is typically suppressed in ultrathin films.\cite{fong2006stabilization}

Another interesting scenario in this system is the relation between spin orientation and FE switching~\cite{di2012electric,picozzi2014ferroelectric,kim2014switchable}. Supposing that the Bloch wave functions of two states with opposite FE polarizations are $|+P, \mathbf{k}\rangle$ and $|-P, \mathbf{k}'\rangle$, respectively, where $P$ denotes the FE polarization, the directions of both $\mathbf{P}$ and $\mathbf{k}$ are reversed under the spatial inversion operation $I$ , i.e., $I|+P, \mathbf{k}\rangle = |-P, -\mathbf{k}\rangle$. The time-reversal operation ${T}$, however, reverses only the wave vector direction, while the polarization remains unchanged. Thus, we have \[ {T} {I}|+P, \mathbf{k}\rangle = |-P, \mathbf{k}\rangle.
\].  The expectation value of the spin operator $\mathbf{S}$ can be written as~\cite{kim2014switchable}:
\begin{align*}
\langle \mathbf{S} \rangle[{-P},{k}] &= \langle -P, \mathbf{k}|\mathbf{S}|-P, \mathbf{k} \rangle \\
&= \langle +P, \mathbf{k} | {I}^{-1} {T}^{-1} \mathbf{S} {T} {I} | +P, \mathbf{k} \rangle \\
&= \langle +P, \mathbf{k} | -\mathbf{S} | +P, \mathbf{k} \rangle = -\langle \mathbf{S} \rangle[{+P}, \mathbf{k}].
\end{align*}

More interestingly, when the direction of FE polarization is reversed by an external electric field, the out-of-plane spin polarization of both inner and outer branches is reversed simultaneously, as illustrated in Figure~\ref{fig:neb}(c). This reveals a fully reversible spin texture governed by FE switching, providing electrical control of spin orientation. Such a property makes (PA)$_2$CsSn$_2$Br$_7$ a compelling all-semiconductor platform for spintronic applications, such as spin field-effect transistors, where fully electric tunability of PSH is highly desirable.

In summary, we have carried out a comprehensive first-principles investigation of the structural, electronic, and spintronic properties of the 2D hybrid perovskite \((\mathrm{PA})_2\mathrm{CsSn}_2\mathrm{Br}_7\). The material exhibits robust ferroelectric polarization along the in-plane $z$-axis, primarily arising from the orientational ordering of the organic cations \(\mathrm{PA}^+\)  and symmetry-breaking distortions in the inorganic \(\mathrm{SnBr}_6\) octahedra.

Our DFT calculations reveal that the FE distortion not only stabilizes spontaneous polarization but also modifies the electronic band structure and stabilizes the bandgap. Incorporation of SOC uncovers significant spin splitting near the band edges, especially in the conduction band with dominant Sn orbital character.

Symmetry analysis confirms that spin splitting is directionally dependent and tied to the underlying crystal symmetries. In particular, we observe that the spin textures are sensitive to the direction of the FE polarization, enabling the reversal of spin orientations through polarization switching--an intrinsic coupling between ferroelectric polarization and spin texture.

These results highlight \((\mathrm{PA})_2\mathrm{CsSn}_2\mathrm{Br}_7\) as a promising candidate for 2D FE semiconductors where electric-field control of spin states could be realized. The combination of SOC-induced spin splitting, switchable ferroelectricity, and a wide bandgap makes this material attractive for spintronic and non-volatile memory applications.

\section{Computational Methods}

All first-principles calculations were performed using the Vienna \textit{Ab initio} Simulation Package (VASP) within the framework of DFT.\cite{kresse1993ab,kresse1996efficient} The projector augmented-wave (PAW) method was used to treat core--valence interactions.\cite{blochl1994projector,kresse1999ultrasoft} The exchange--correlation potential was described using the Perdew-BurkeErnzerhof functional revised for solids (PBEsol), chosen for its balanced accuracy in predicting structural and electronic properties.\cite{pbesol} The plane-wave cutoff was set to 570 eV. Brillouin-zone integrations employed $\Gamma$-centered meshes of $4 \times 4 \times 4$ for bulk and $1 \times 4 \times 4$ for monolayer systems; convergence tests ensured that total energies varied by less than 1 meV/atom upon further mesh refinement.  

For bulk calculations, lattice constants were fixed to the optimized experimental values, while atomic positions were fully relaxed. The convergence thresholds were $10^{-6}$ eV for total energy and $0.001$ eV/\AA{} for Hellmann--Feynman forces. A Gaussian smearing of 0.05 eV was used for all calculations.\cite{elsasser1994density} In monolayer systems, a vacuum region of 20 \AA{} was introduced along the non-periodic direction to suppress spurious interlayer interactions.  

Noncollinear spin-polarized calculations with SOC were carried out for both bulk and monolayer structures. FE switching pathways were evaluated using the nudged elastic band (NEB) method,\cite{henkelman2000climbing,absor2019intrinsic,ai2019reversible} while spontaneous polarization was computed within the Berry-phase formalism of the modern theory of polarization.\cite{spaldin2012beginner,resta1994macroscopic,king1993theory} Both polar and centrosymmetric reference structures were considered to estimate switchable polarization.  

Spin textures were obtained from DFT using a dense $10 \times 10$ \textbf{k}-mesh around the $\Gamma$ point in the $k_y$--$k_z$ plane. Model Hamiltonians were parametrized by fitting to the DFT band dispersions via minimization of  
\begin{equation}
S = \sum_{i=1}^{2} \sum_{\mathbf{k}} f(\mathbf{k}) 
\left| \det\!\left[ H(\mathbf{k}) - E^i(\mathbf{k}) I \right] \right|^2,
\end{equation}
where $E^i(\mathbf{k})$ are the target DFT eigenvalues, $H(\mathbf{k})$ is the model Hamiltonian, and $f(\mathbf{k})$ is a Gaussian weight centered at $\Gamma$ to emphasize accuracy near the band extrema.\cite{ai2019reversible,sasmito2021reversible}  

Symmetry analyses were conducted using the Bilbao Crystallographic Server,\cite{aroyo2006krou} SEEK-PATH,\cite{hinuma2017band} and FINDSYM.\cite{stokes2005findsym} Spin textures were visualized using PYPROCAR,\cite{herath2020pyprocar}  and band structures were visualized using VESTA\cite{momma2011vesta} while symbolic manipulations were performed in MATHEMATICA.\cite{wolfram}

\section*{Supporting Information}

The Supporting Information provides additional results and analyses supporting this study. It includes the crystal structures of various configurations of \((\mathrm{PA})_2\mathrm{CsY}_2\mathrm{X}_7\) (Y = Sn, Pb; X = Br, I), and the electronic band structures of (PA)$_2$CsPb$_2$Br$_7$, (PA)$_2$CsPb$_2$I$_7$, and (PA)$_2$CsSn$_2$I$_7$. Furthermore, detailed DFT and $k\cdot{p}$ spin texture analyses near the $\Gamma$ point are presented for bulk and monolayer (PA)$_2$CsPb$_2$Br$_7$, (PA)$_2$CsPb$_2$I$_7$, and (PA)$_2$CsSn$_2$I$_7$. Finally, polarization switching and the resulting reversible spin textures in monolayer (PA)$_2$CsSn$_2$Br$_7$ are discussed.

\section{Acknowledgement}
D.T. acknowledges UGC, India, for the junior research fellowship [Grant No. BUPME01420733/(CSIR-UGC-NET JUNE 2022)]. S.B. acknowledges financial support from SERB under a core research grant (Grant No. CRG/2019/000647) to set up his High Performance Computing (HPC) facility ``Veena'' at IIT Delhi for computational resources.

\bibliography{reference.bib}

\end{document}


\begin{center}
    \Large \textbf{Ferroelectric Control of Spin Textures in Layered Hybrid Perovskites} \\[1em]

    \normalsize
    Divyanshi Tyagi$^{*}$, Saswata Bhattacharya$^{\dagger}$ \\[0.5em]
    \textit{Department of Physics, Indian Institute of Technology Delhi, New Delhi 110016, India} \\[2em]

    \LARGE \textbf{Supplementary Information}
\end{center}

\vspace{2em}

\noindent \textbf{I.} Crystal structures of various configurations of \((\mathrm{PA})_2\mathrm{CsY}_2\mathrm{X}_7\);Y=Sn,Pb and X= Br,I. \\

\noindent \textbf{II.} The electronic band structures of (PA)$_2$CsPb$_2$Br$_7$, (PA)$_2$CsPb$_2$I$_7$, and (PA)$_2$CsSn$_2$I$_7$. \\

\noindent \textbf{III.} Bulk (PA)$_2$CsPb$_2$Br$_7$: DFT and $\boldsymbol{k \cdot p}$ spin textures near $\Gamma$ \\

\noindent \textbf{IV.}  Monolayer (PA)$_2$CsPb$_2$Br$_7$: DFT and $\boldsymbol{k \cdot p}$ spin textures near $\Gamma$ \\

\noindent \textbf{V.}  Bulk (PA)$_2$CsPb$_2$I$_7$: DFT and $\boldsymbol{k \cdot p}$ spin textures near $\Gamma$ \\

\noindent \textbf{VI.}  Monolayer (PA)$_2$CsPb$_2$I$_7$: DFT and $\boldsymbol{k \cdot p}$ spin textures near $\Gamma$. \\

\noindent \textbf{VII.} Bulk (PA)$_2$CsSn$_2$I$_7$: DFT and $\boldsymbol{k \cdot p}$ spin textures near $\Gamma$ \\

\noindent \textbf{VIII.} Monolayer (PA)$_2$CsSn$_2$I$_7$: DFT and $\boldsymbol{k \cdot p}$ spin textures near $\Gamma$ \\

\noindent \textbf{IX.} Polarization switching and reversible spin textures in monolayer (PA)$_2$CsSn$_2$Br$_7$ \\

\vfill

\noindent $^{*}$ divyanshi.tyagi@physics.iitd.ac.in \\
\noindent $^{\dagger}$  saswata@physics.iitd.ac.in

\newpage
\section*{I. Crystal structures of various configurations of \((\mathrm{PA})_2\mathrm{CsY}_2\mathrm{X}_7\);Y=Sn,Pb and X= Br,I.}
\begin{figure}[h!]
\centering
\includegraphics[width=0.7\textwidth]{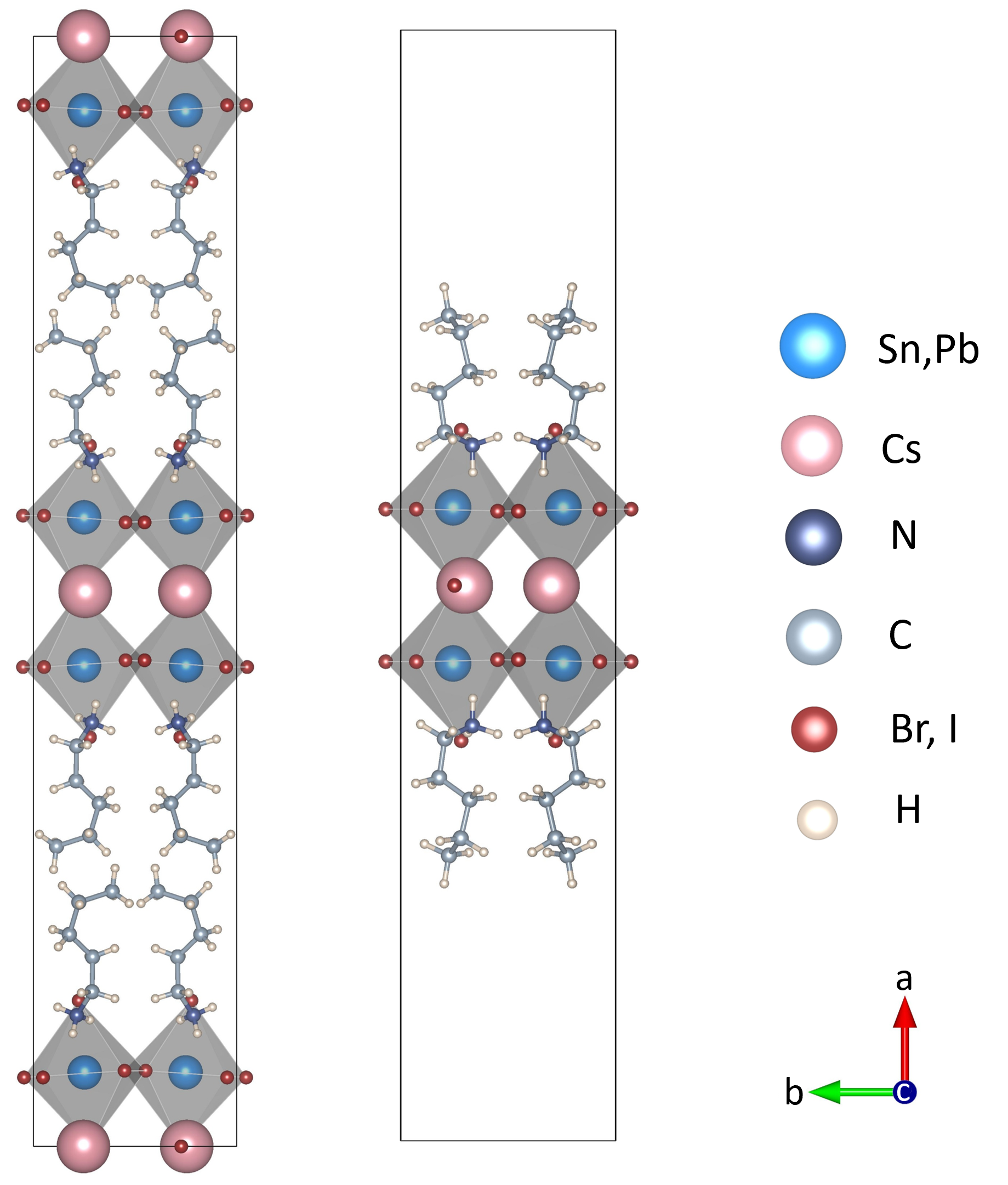} 
\caption{Side view of the relaxed crystal structures of bulk (left) and monolayer (right) (PA)$_2$CsY$_2$X$_7$ : Y= Pb,Sn and X= I, Br}
\end{figure}

\newpage
\section*{II. The electronic band structures of  (PA)$_2$CsPb$_2$Br$_7$, (PA)$_2$CsPb$_2$I$_7$, and (PA)$_2$CsSn$_2$I$_7$}
\begin{figure}[h!]
\centering
\includegraphics[width=0.9\textwidth]{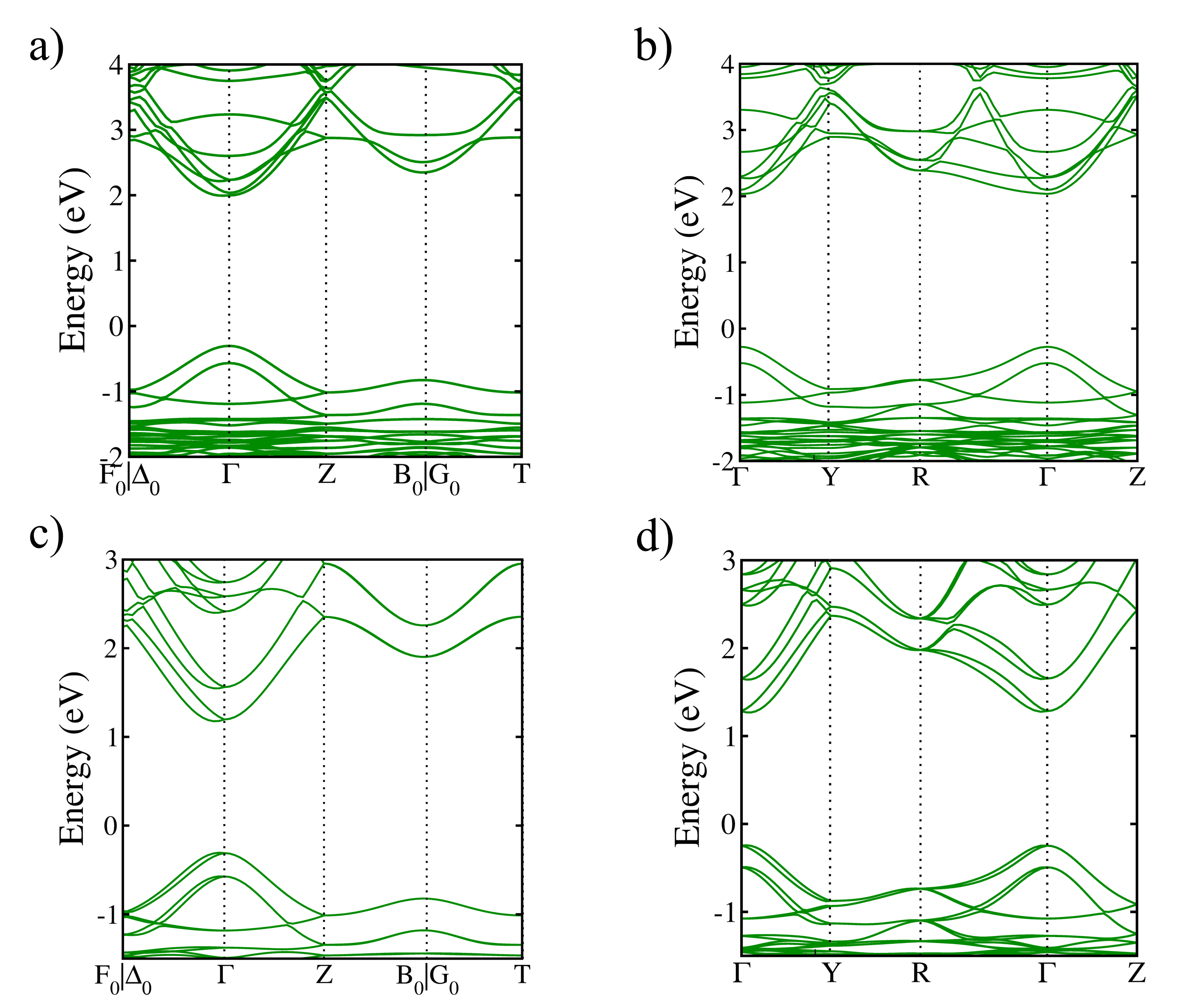} 
\caption{Electronic band structures of bulk (a, c) and monolayer (b, d) (PA)$_2$CsPb$_2$Br$_7$. 
Panels (a, b) show results without spin--orbit coupling (SOC), while (c, d) include SOC.}
\end{figure}

\begin{figure}[!b]
\centering
\includegraphics[width=0.9\textwidth]{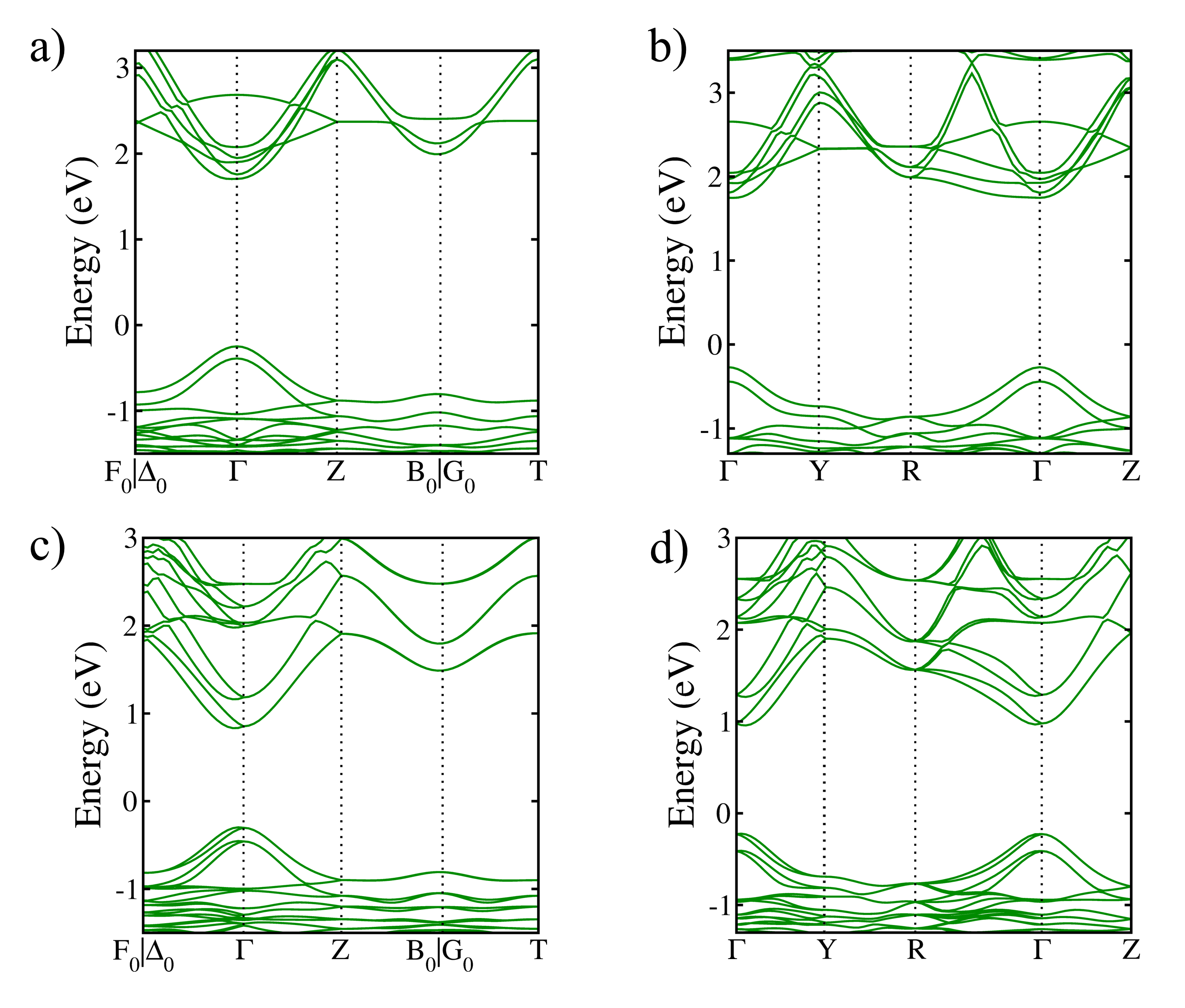} 
\caption{(a) Electronic band structures of bulk (a, c) and monolayer (b, d) (PA)$_2$CsPb$_2$I$_7$. 
Panels (a, b) show results without spin--orbit coupling (SOC), while (c, d) include SOC.}
\end{figure}

\newpage

\begin{figure}
\centering
\includegraphics[width=0.9\textwidth]{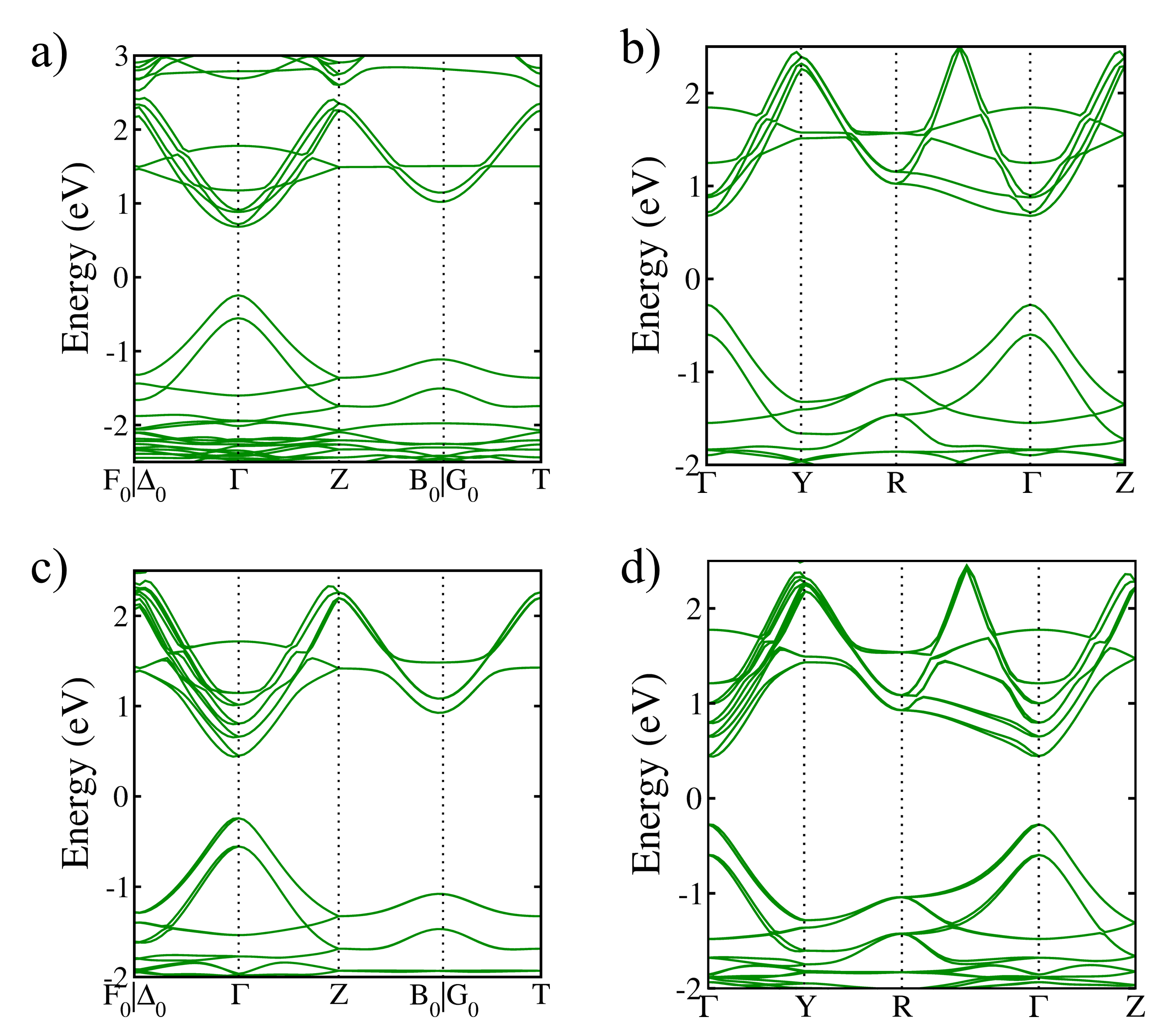} 
\caption{Electronic band structures of bulk (a, c) and monolayer (b, d) (PA)$_2$CsSn$_2$I$_7$. 
Panels (a, b) show results without spin--orbit coupling (SOC), while (c, d) include SOC.}
\end{figure}

\clearpage

\section*{III. Bulk \((\mathrm{PA})_2\mathrm{CsPb}_2\mathrm{Br}_7\): spin textures and $\boldsymbol{k \cdot p}$ model near $\Gamma$.}

\begin{figure}[h!]
\centering
\includegraphics[width=1.1\textwidth]{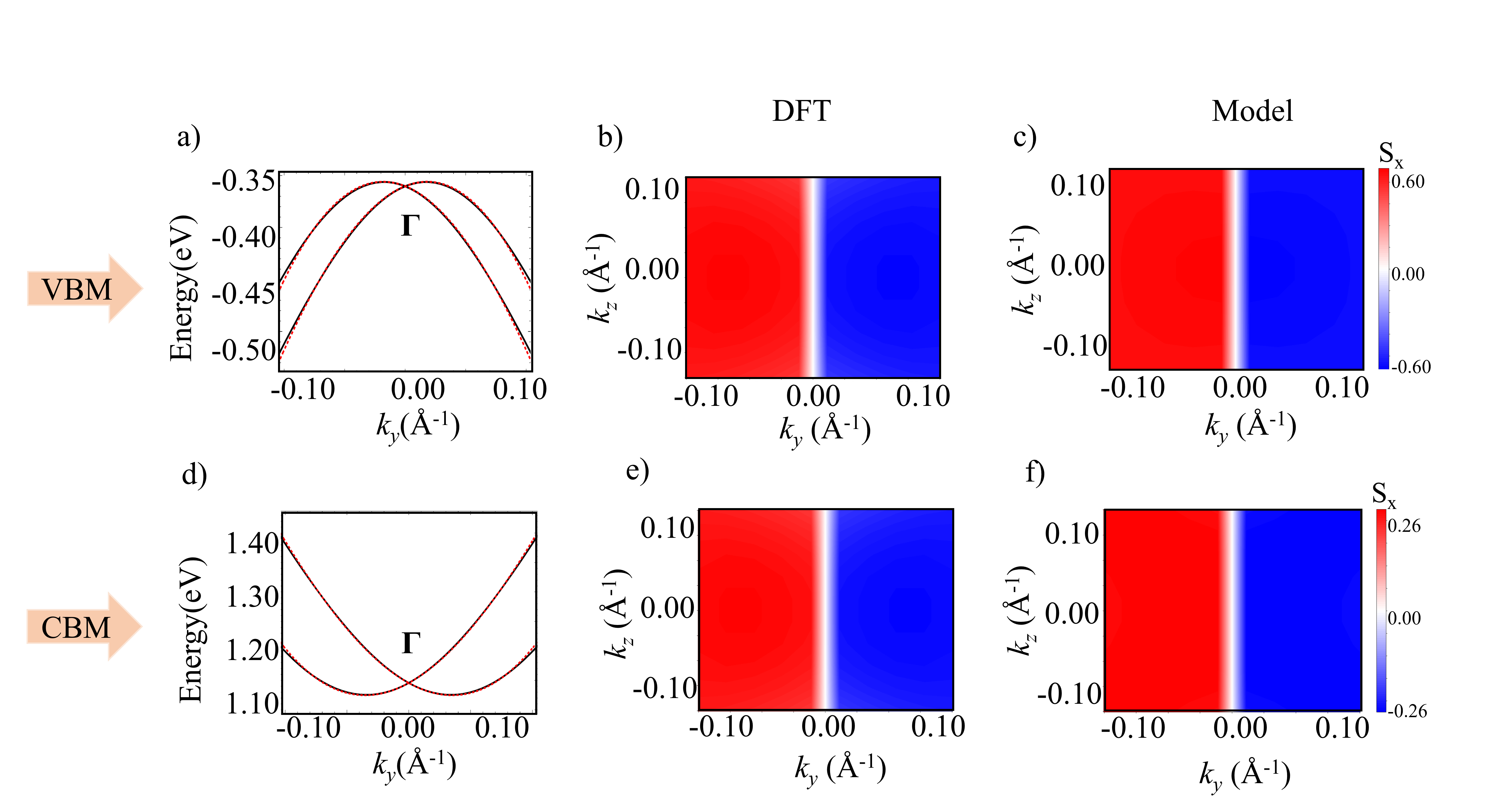} 
\caption{Bulk (PA)$_2$CsPb$_2$Br$_7$: out-of-plane spin textures near $\Gamma$. 
Top row = valence band maxima (VBM), bottom row = conduction band minima (CBM). 
(a,d) DFT band dispersion along $k_y$ (black) with $\boldsymbol{k \cdot p}$ fits (red dashed); 
(b,e) DFT $S_x$ on constant-energy contours; 
(c,f) $S_x$ from the two-band $\boldsymbol{k \cdot p}$ model. 
Axes in \AA$^{-1}$. Colorbar ranges differ between rows.}
\label{fig:bulk_pst}
\end{figure}

\begin{table}[!b]
\centering
\caption{Fitted $\beta^{(1)}$ and $\delta$ coefficients and the spin splitting from DFT for the VBM and CBM in bulk (PA)$_2$CsPb$_2$Br$_7$. Units: $\beta^{(1)}$ in eV\,$\cdot$\,\AA, $\delta$ in eV\,$\cdot$\,\AA$^3$, splitting in eV.}
\label{tab:bulk_params}
\begin{tabular}{
    l
    S[table-format=1.2]
    S[table-format=3.2]
    S[table-format=1.2]
}
\toprule
\textbf{Band} & {$\beta^{(1)}$} & {$\delta$} & {Splitting (DFT)} \\
\midrule
VBM & 0.46 & -12.52 & 0.44 \\
CBM & 1.15 & -23.39 & 1.27 \\
\bottomrule
\end{tabular}
\end{table}

\newpage
\section*{IV. Monolayer (PA)$_2$CsPb$_2$Br$_7$: DFT and $\boldsymbol{k \cdot p}$ spin textures near $\Gamma$}

\begin{figure}[h!]
\centering
\includegraphics[width=1.1\textwidth]{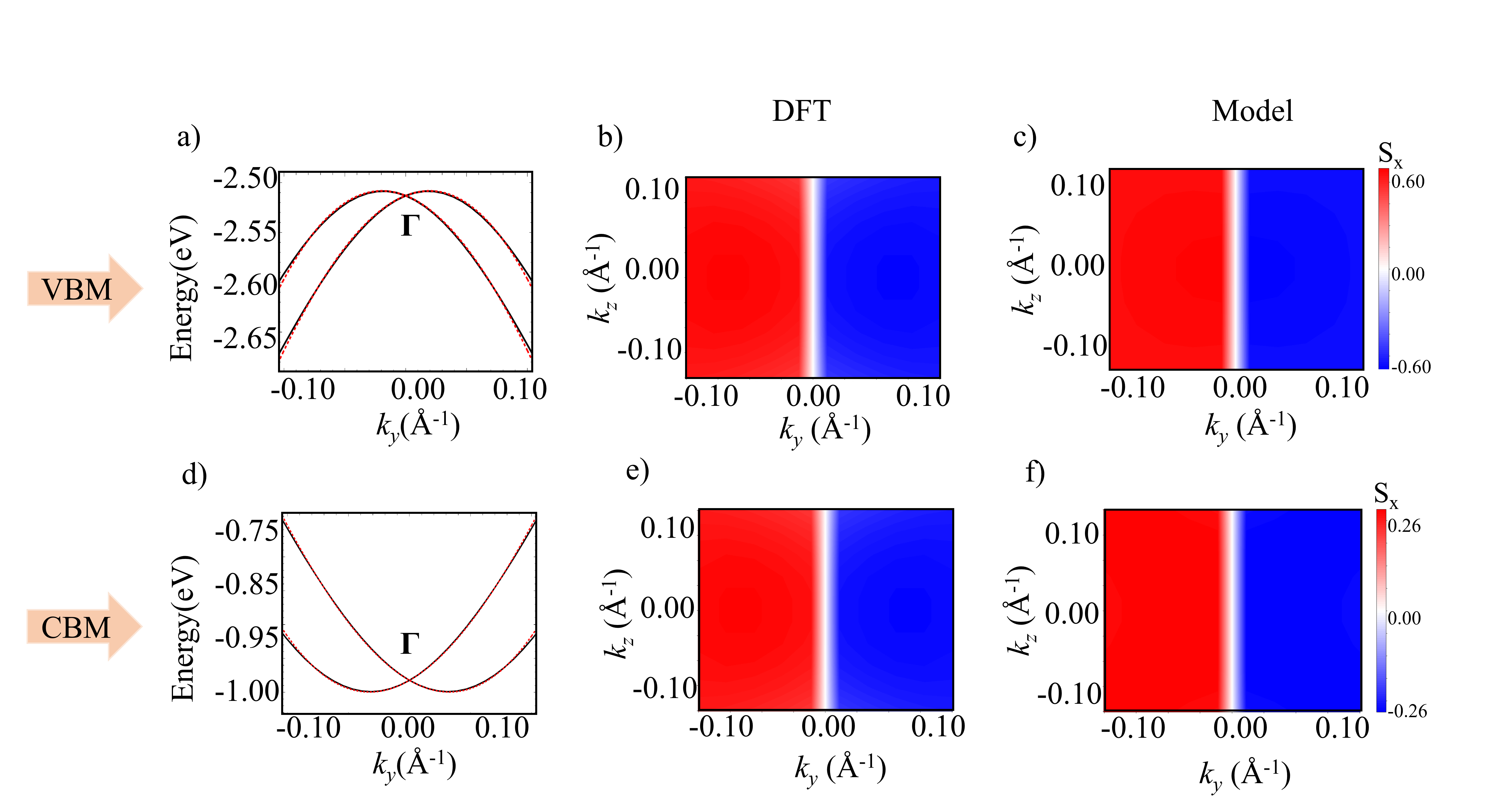} \hspace{1cm} 
\caption{Monolayer (PA)$_2$CsPb$_2$Br$_7$: out-of-plane spin textures near $\Gamma$. 
Top row = VBM, bottom row = CBM. 
(a,d) DFT band dispersion along $k_y$ (black) with $\boldsymbol{k \cdot p}$ fits (red dashed); 
(b,e) DFT $S_x$ on constant-energy contours; 
(c,f) $S_x$ from the two-band $\boldsymbol{k \cdot p}$ model. 
Axes in \AA$^{-1}$. Colorbar ranges differ between rows.}

\end{figure}

\begin{table}[!b]
\centering
\caption{Fitted $\beta^{(1)}$ and $\delta$ coefficients and the spin splitting from DFT for the VBM and CBM in monolayer (PA)$_2$CsPb$_2$Br$_7$. Units: $\beta^{(1)}$ in eV\,$\cdot$\,\AA, $\delta$ in eV\,$\cdot$\,\AA$^3$, splitting in eV.}
\label{tab:mono_params}
\begin{tabular}{
    l
    S[table-format=1.2]
    S[table-format=3.2]
    S[table-format=1.2]
}
\toprule
\textbf{Band} & {$\beta^{(1)}$} & {$\delta$} & {Splitting (DFT)} \\
\midrule
VBM & 0.48 & -14.75 & 0.58 \\
CBM & 1.03 & -21.20 & 1.08 \\
\bottomrule
\end{tabular}
\end{table}

\newpage
\section*{V. Bulk (PA)$_2$CsPb$_2$I$_7$: DFT and $\boldsymbol{k \cdot p}$ spin textures near $\Gamma$}

\begin{figure}[h!]
\centering
\includegraphics[width=1.1\textwidth]{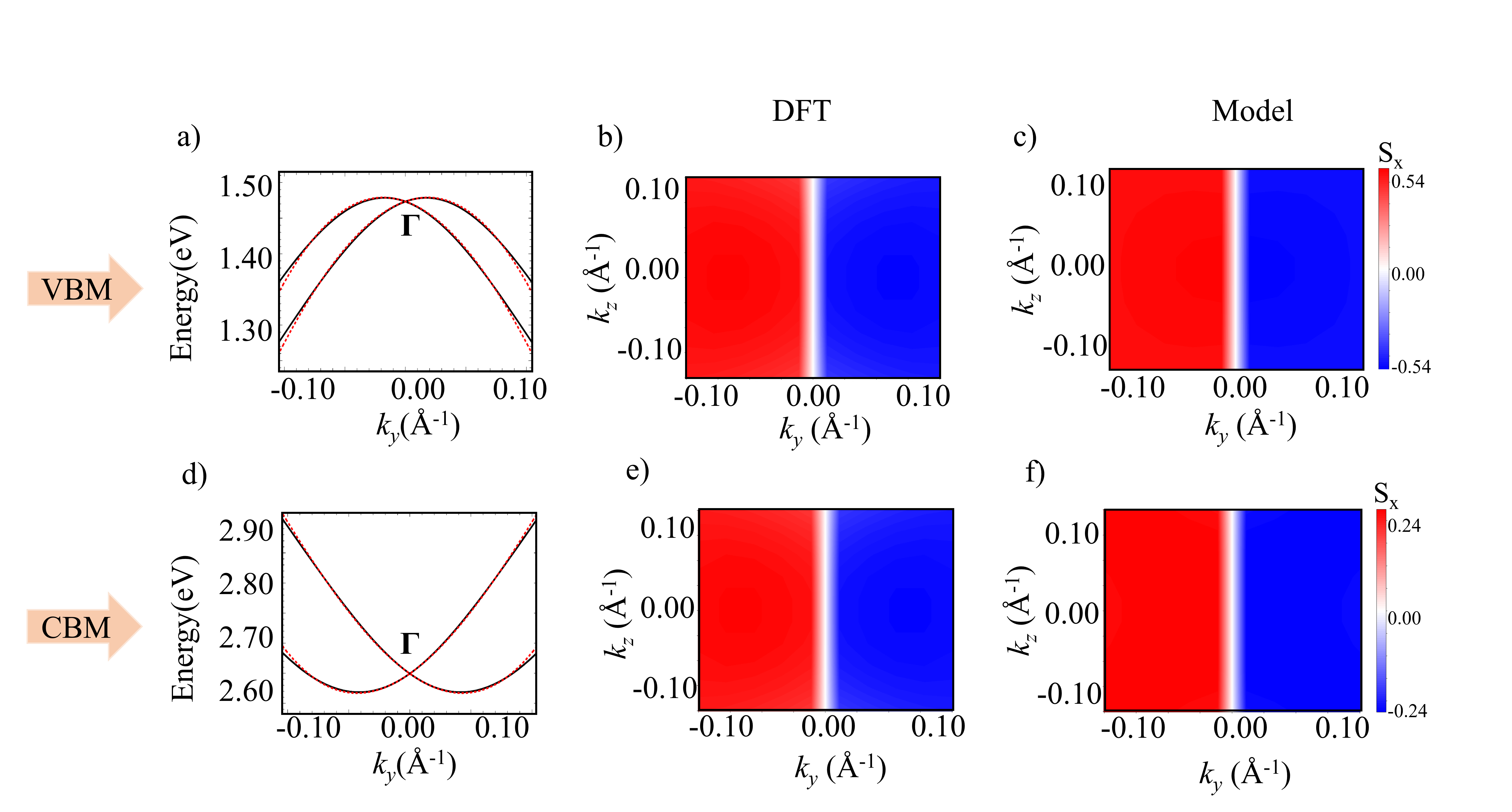} \hspace{1cm} 
\caption{Bulk (PA)$_2$CsPb$_2$I$_7$: out-of-plane spin textures near $\Gamma$. 
Top row = VBM, bottom row = CBM. 
(a,d) DFT band dispersion along $k_y$ (black) with $\boldsymbol{k \cdot p}$ fits (red dashed); 
(b,e) DFT $S_x$ on constant-energy contours; 
(c,f) $S_x$ from the two-band $\boldsymbol{k \cdot p}$ model. 
Axes in \AA$^{-1}$. Colorbar ranges differ between rows.}

\end{figure}

\begin{table}[!b]
\centering
\caption{Fitted $\beta^{(1)}$ and $\delta$ coefficients and the spin splitting from DFT for the VBM and CBM in bulk (PA)$_2$CsPb$_2$I$_7$. Units: $\beta^{(1)}$ in eV\,$\cdot$\,\AA, $\delta$ in eV\,$\cdot$\,\AA$^3$, splitting in eV.}
\label{tab:bulk_params}
\begin{tabular}{
    l
    S[table-format=1.2]
    S[table-format=3.2]
    S[table-format=1.2]
}
\toprule
\textbf{Band} & {$\beta^{(1)}$} & {$\delta$} & {Splitting (DFT)} \\
\midrule
VBM & 0.59 & -16.65 & 0.68 \\
CBM & 1.43 & -34.25 & 1.57 \\
\bottomrule
\end{tabular}
\end{table}

\newpage
\section*{VI. Monolayer (PA)$_2$CsPb$_2$I$_7$: DFT and $\boldsymbol{k \cdot p}$ spin textures near $\Gamma$}

\begin{figure}[h!]
\centering
\includegraphics[width=1.1\textwidth]{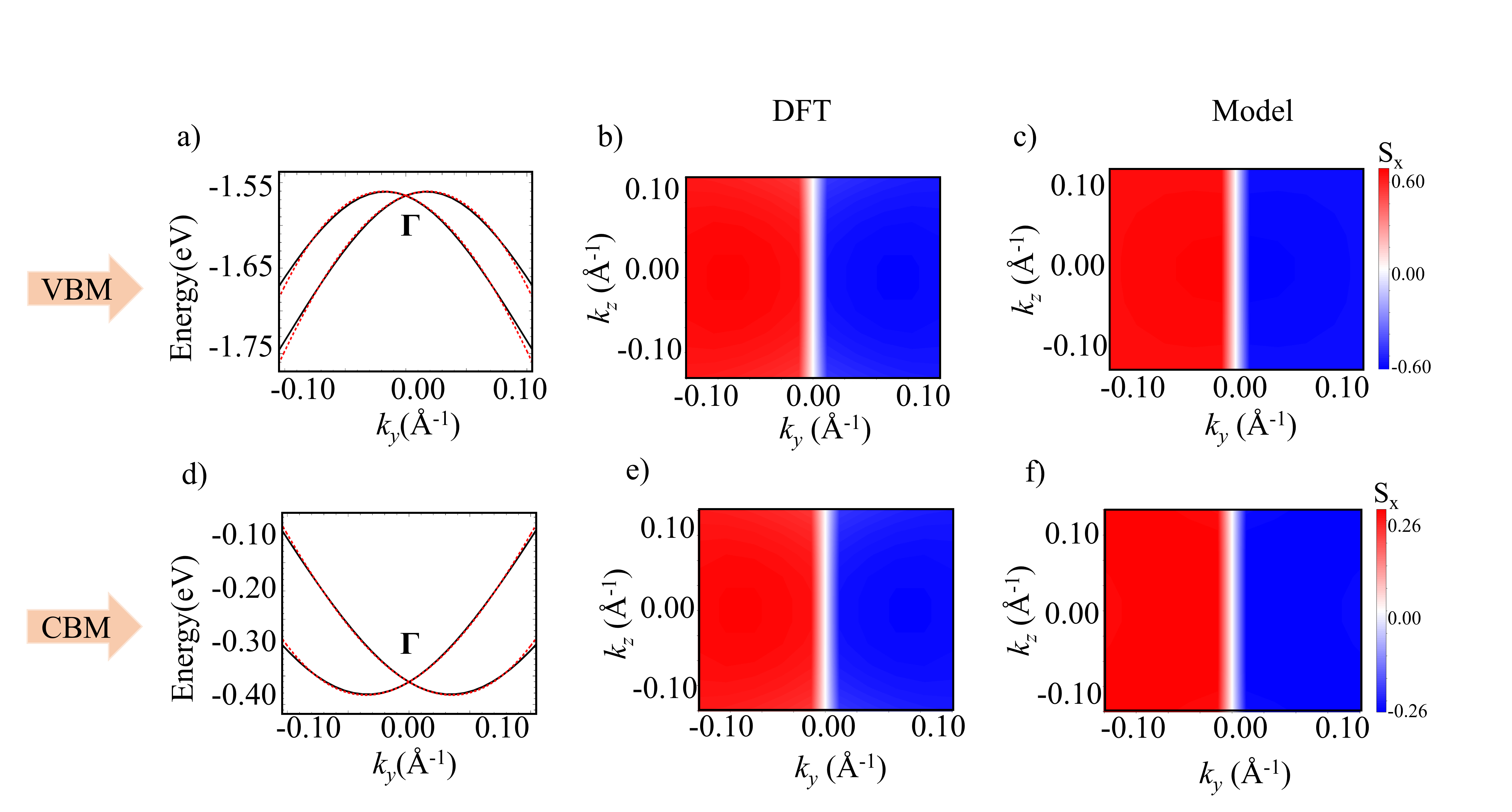} \hspace{1cm} 
\caption{Bulk (PA)$_2$CsPb$_2$I$_7$: out-of-plane spin textures near $\Gamma$. 
Top row = VBM, bottom row = CBM. 
(a,d) DFT band dispersion along $k_y$ (black) with $\boldsymbol{k \cdot p}$ fits (red dashed); 
(b,e) DFT $S_x$ on constant-energy contours; 
(c,f) $S_x$ from the two-band $\boldsymbol{k \cdot p}$ model. 
Axes in \AA$^{-1}$. Colorbar ranges differ between rows.}

\end{figure}

\begin{table}[!b]
\centering
\caption{Fitted $\beta^{(1)}$ and $\delta$ coefficients and the spin splitting from DFT for the VBM and CBM in monolayer (PA)$_2$CsPb$_2$I$_7$. Units: $\beta^{(1)}$ in eV\,$\cdot$\,\AA, $\delta$ in eV\,$\cdot$\,\AA$^3$, splitting in eV.}
\label{tab:mono_iodi_params}
\begin{tabular}{
    l
    S[table-format=1.2]
    S[table-format=3.2]
    S[table-format=1.2]
}
\toprule
\textbf{Band} & {$\beta^{(1)}$} & {$\delta$} & {Splitting (DFT)} \\
\midrule
VBM & 0.54 & -20.25 & 0.61 \\
CBM & 1.25 & -26.85 & 1.43 \\
\bottomrule
\end{tabular}
\end{table}

\newpage
\section*{VII. Bulk (PA)$_2$CsSn$_2$I$_7$: DFT and $\boldsymbol{k \cdot p}$ spin textures near $\Gamma$}

\begin{figure}[h!]
\centering
\includegraphics[width=1.1\textwidth]{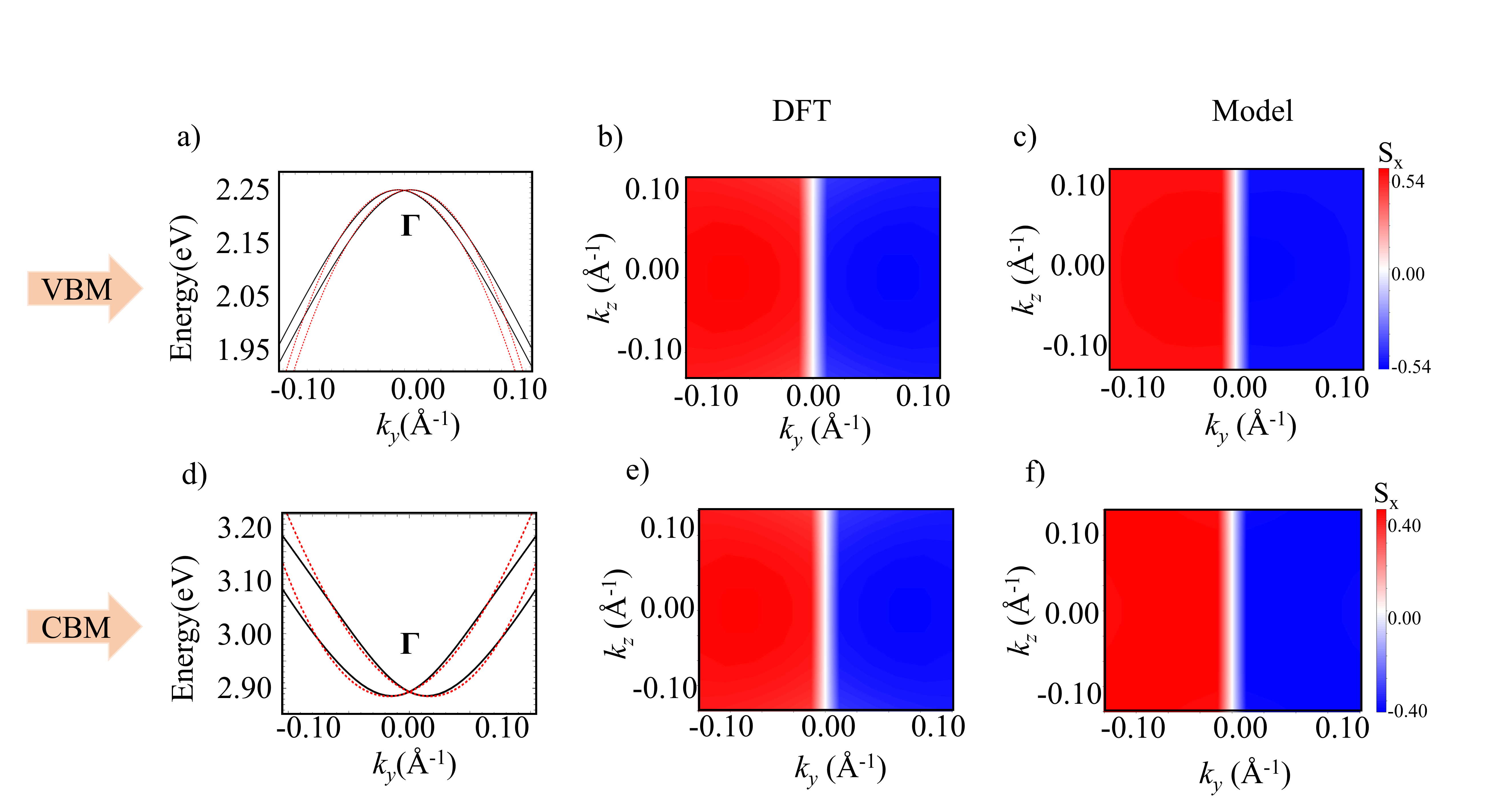} \hspace{1cm} 
\caption{Bulk (PA)$_2$CsPb$_2$I$_7$: out-of-plane spin textures near $\Gamma$. 
Top row = VBM, bottom row = CBM. 
(a,d) DFT band dispersion along $k_y$ (black) with $\boldsymbol{k \cdot p}$ fits (red dashed); 
(b,e) DFT $S_x$ on constant-energy contours; 
(c,f) $S_x$ from the two-band $\boldsymbol{k \cdot p}$ model. 
Axes in \AA$^{-1}$. Colorbar ranges differ between rows.}

\end{figure}

\begin{table}[!b]
\centering
\caption{Fitted $\beta^{(1)}$ and $\delta$ coefficients and the spin splitting from DFT for the VBM and CBM in bulk (PA)$_2$CsSn$_2$I$_7$. Units: $\beta^{(1)}$ in eV\,$\cdot$\,\AA, $\delta$ in eV\,$\cdot$\,\AA$^3$, splitting in eV.}
\label{tab:bulk_sni_params}
\begin{tabular}{
    l
    S[table-format=1.2]
    S[table-format=3.2]
    S[table-format=1.2]
}
\toprule
\textbf{Band} & {$\beta^{(1)}$} & {$\delta$} & {Splitting (DFT)} \\
\midrule
VBM & 0.41 & -17.63 & 0.49 \\
CBM & 0.94 & -38.66 & 1.03 \\
\bottomrule
\end{tabular}
\end{table}

\newpage
\section*{VIII. Monolayer (PA)$_2$CsSn$_2$I$_7$: DFT and $\boldsymbol{k \cdot p}$ spin textures near $\Gamma$}

\begin{figure}[h!]
\centering
\includegraphics[width=1.1\textwidth]{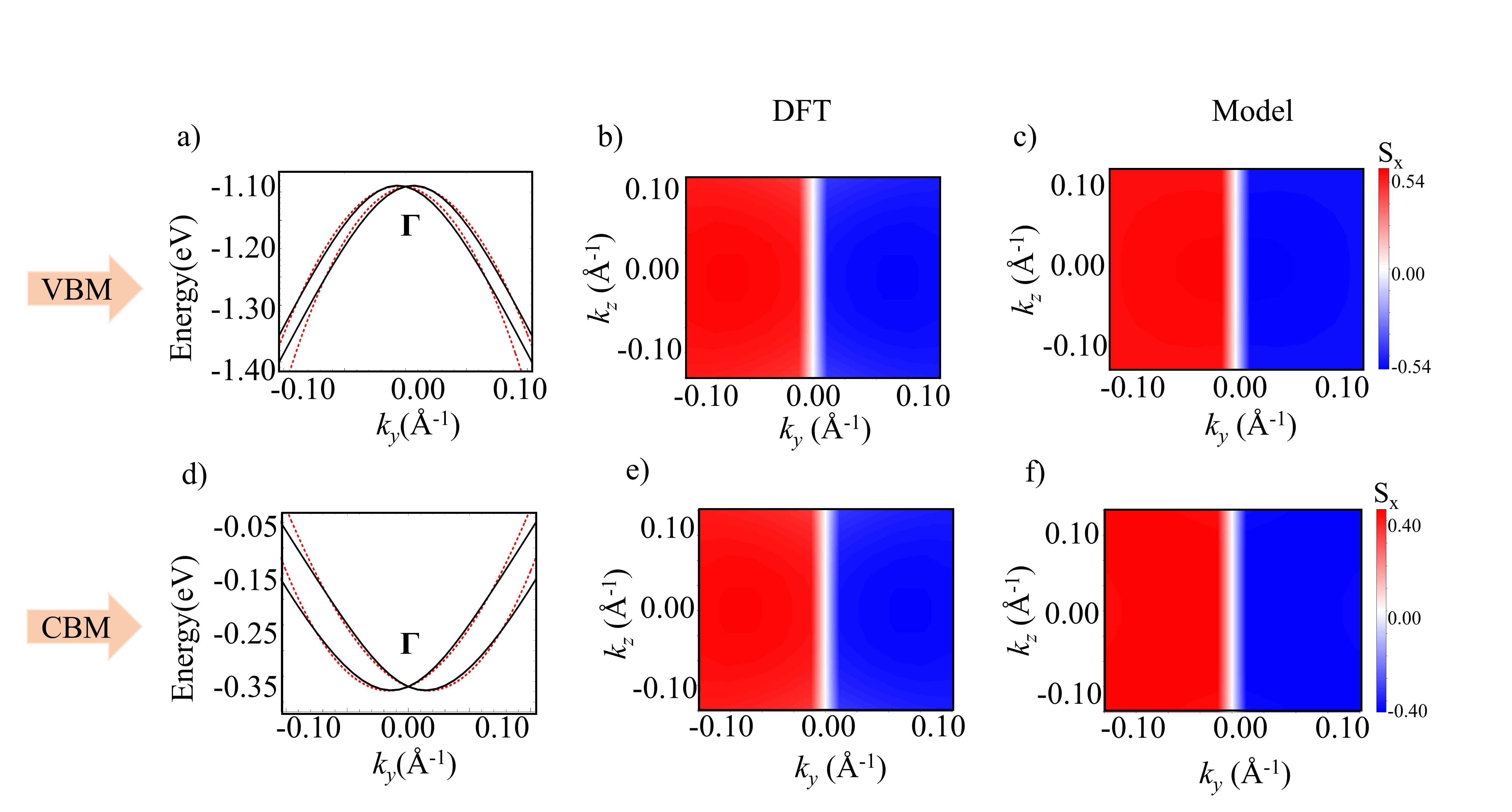} \hspace{1cm} 
\caption{Bulk (PA)$_2$CsSn$_2$I$_7$: out-of-plane spin textures near $\Gamma$. 
Top row = VBM, bottom row = CBM. 
(a,d) DFT band dispersion along $k_y$ (black) with $\boldsymbol{k \cdot p}$ fits (red dashed); 
(b,e) DFT $S_x$ on constant-energy contours; 
(c,f) $S_x$ from the two-band $\boldsymbol{k \cdot p}$ model. 
Axes in \AA$^{-1}$. Colorbar ranges differ between rows.}

\end{figure}

\begin{table}[!b]
\centering
\caption{Fitted $\beta^{(1)}$ and $\delta$ coefficients and the spin splitting from DFT for the VBM and CBM in monolayer (PA)$_2$CsSn$_2$I$_7$. Units: $\beta^{(1)}$ in eV\,$\cdot$\,\AA, $\delta$ in eV\,$\cdot$\,\AA$^3$, splitting in eV.}
\label{tab:mono_sni_params}
\begin{tabular}{
    l
    S[table-format=1.2]
    S[table-format=3.2]
    S[table-format=1.2]
}
\toprule
\textbf{Band} & {$\beta^{(1)}$} & {$\delta$} & {Splitting (DFT)} \\
\midrule
VBM & 0.32 & 17.80 & 0.42 \\
CBM & 0.85 & -17.04 & 0.96 \\
\bottomrule
\end{tabular}
\end{table}

\newpage
\section*{IX. Polarization switching and reversible spin textures in monolayer (PA)$_2$CsSn$_2$Br$_7$}
\begin{figure}[h!]
\centering
\includegraphics[width=0.9\textwidth]{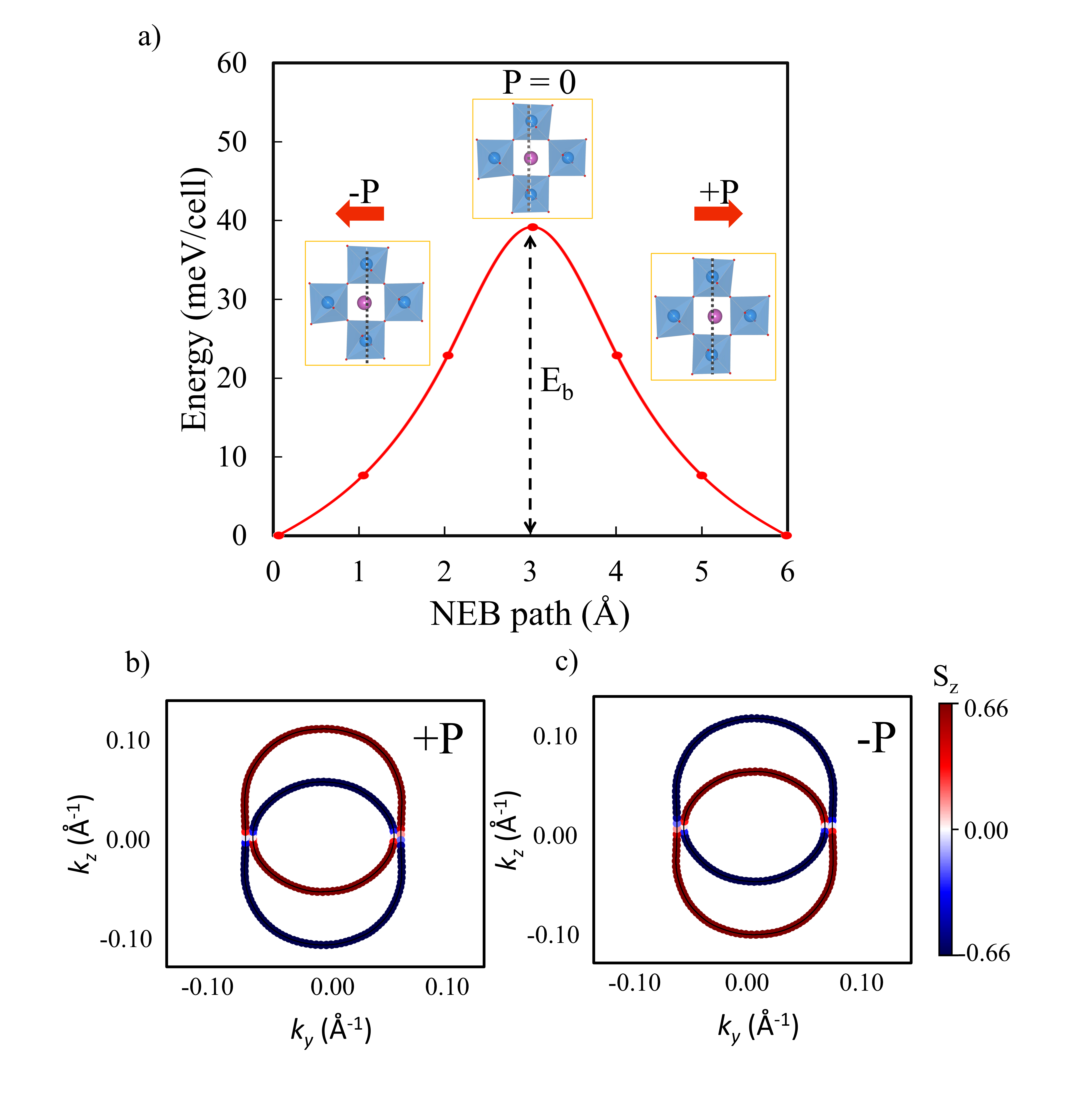} 
\caption{(a) Climbing‐-image nudged elastic band (CI‐-NEB) calculation for the polarization switching process in monolayer (PA)$_2$CsSn$_2$Br$_7$ perovskite. Two ferroelectric structures in the ground state with opposite directions of electric polarization are shown. $E_b$ is the activation barrier energy for the polarization switching process which is 40meV for monolayer. Reversible in‐plane spin textures calculated at constant energy $E = E_F + 1.5$eV with opposite spin polarization: (b) $-P$, (c) $+P$.%
.}
\end{figure}